\documentclass[letterpaper]{article} 
\usepackage[submission]{aaai23}  
\usepackage{times}  
\usepackage{helvet}  
\usepackage{courier}  
\usepackage[hyphens]{url}  
\usepackage{graphicx} 
\usepackage{natbib}  
\usepackage{caption} 
\usepackage{algorithm}
\usepackage{algorithmic}

\usepackage{amsmath}
\usepackage{amsfonts}
\usepackage{graphicx}
\usepackage{subfigure}
\usepackage{multirow}

\usepackage{newfloat}
\usepackage{listings}
\DeclareCaptionStyle{ruled}{labelfont=normalfont,labelsep=colon,strut=off} 
\floatstyle{ruled}
\newfloat{listing}{tb}{lst}{}
\floatname{listing}{Listing}
\title{HAVEN: Hierarchical Cooperative Multi-Agent Reinforcement Learning\\ with Dual Coordination Mechanism}
\author {
    Zhiwei Xu,
    Yunpeng Bai,
    Bin Zhang,
    Dapeng Li,
    Guoliang Fan
}
\affiliations {
    Institute of Automation, Chinese Academy of Sciences\\
    School of Artificial Intelligence, University of Chinese Academy of Sciences\\
    Beijing, China\\
    \{xuzhiwei2019, baiyuanpeng2020, zhangbin2020, lidapeng2020, guoliang.fan\}@ia.ac.cn
}

\begin{document}

\maketitle

\begin{abstract}
Recently, some challenging tasks in multi-agent systems have been solved by some hierarchical reinforcement learning methods. Inspired by the intra-level and inter-level coordination in the human nervous system, we propose a novel value decomposition framework HAVEN based on hierarchical reinforcement learning for fully cooperative multi-agent problems. To address the instability arising from the concurrent optimization of policies between various levels and agents, we introduce the dual coordination mechanism of inter-level and inter-agent strategies by designing reward functions in a two-level hierarchy. HAVEN does not require domain knowledge and pre-training, and can be applied to any value decomposition variant. Our method achieves desirable results on different decentralized partially observable Markov decision process domains and outperforms other popular multi-agent hierarchical reinforcement learning algorithms.
\end{abstract}


\section{Introduction}

There has been a growing interest in multi-agent reinforcement learning (MARL) in the last few years, which plays a vital role in various tasks such as traffic control~\cite{Kuyer2008MultiagentRL} and recommendation systems~\cite{Choi2018ReinforcementLB}. Most MARL algorithms follow the paradigm known as centralized training with decentralized execution (CTDE)~\cite{Lowe2017MultiAgentAF}. Each agent can utilize all available information during training but can only make decisions based on local observations. According to this principle, MARL algorithms can be divided into several categories, including those based on centralized critics and decentralized actors~\cite{Lowe2017MultiAgentAF,Foerster2018CounterfactualMP,Iqbal2019ActorAttentionCriticFM}, communication~\cite{Sukhbaatar2016LearningMC,Foerster2016LearningTC,Peng2017MultiagentBN}, and value decomposition~\cite{Sunehag2018ValueDecompositionNF,Rashid2018QMIXMV,Son2019QTRANLT}. In a fully cooperative scenario, value decomposition methods can significantly alleviate the credit assignment issue. Numerous value decomposition variants with significant performance have been proposed recently.

\begin{figure}[t]
    \centering
    \includegraphics[width=2.2 in]{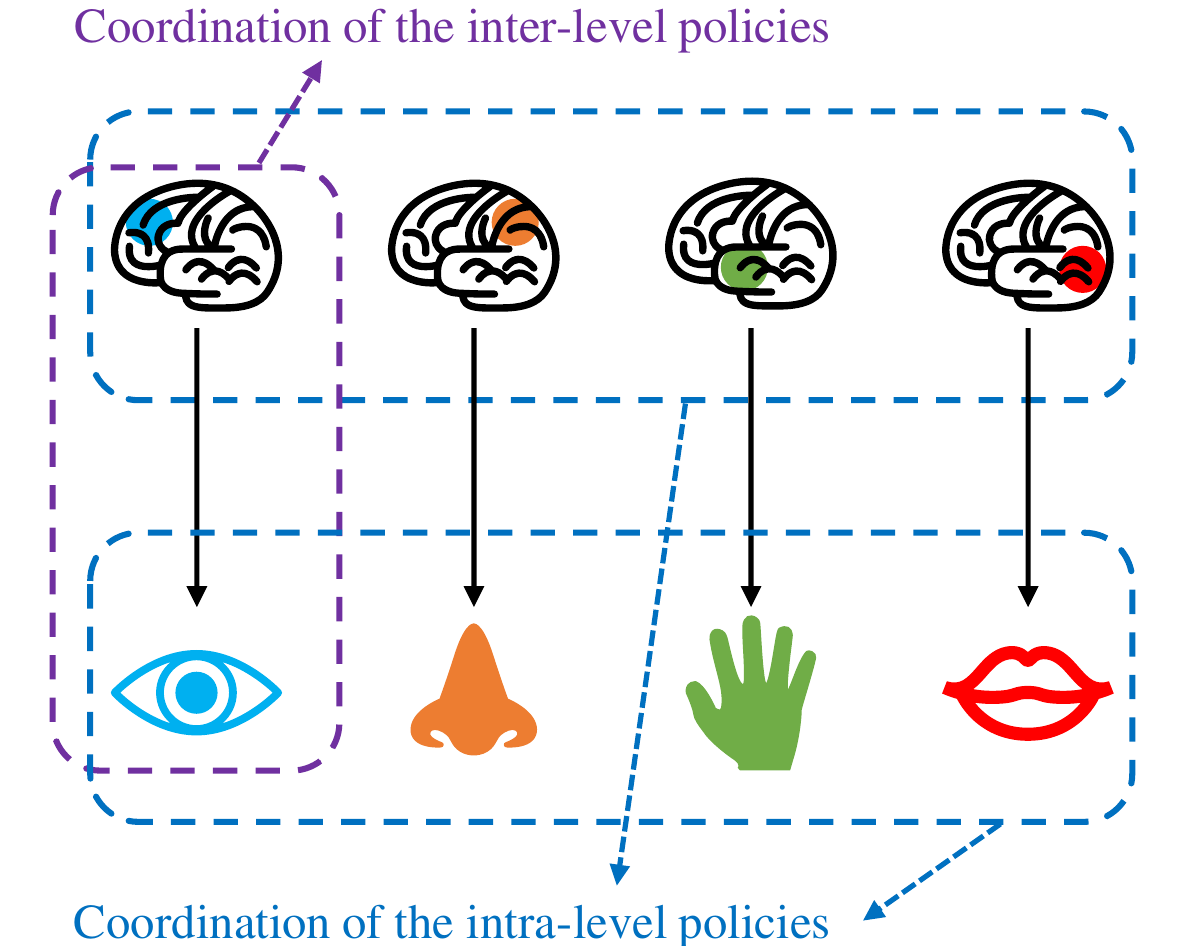}
    \caption{Inter-level and intra-level coordination in the human nervous system. We can consider the upper half of the figure as high-level policies in hierarchical reinforcement learning, and the lower half as low-level policies.}
    \label{fig:intro}
\end{figure}

However, most previous studies on multi-agent cooperative tasks do not consider hierarchical structures. Hierarchical reinforcement learning (HRL) is a computational approach that learns to operate on different levels of temporal abstraction. Traditional HRL methods include hierarchical abstraction machine (HAM)~\cite{Parr1997ReinforcementLW}, MAXQ~\cite{Dietterich2000HierarchicalRL}, option~\cite{Sutton1999BetweenMA,Precup2000TemporalAI}, and feudal architecture~\cite{Dayan1992FeudalRL}. With the emergence of deep learning, HRL has gradually evolved into two distinct branches: subgoal-based~\cite{Vezhnevets2017FeUdalNF,Nachum2018DataEfficientHR} and option-based methods~\cite{Bacon2017TheOA,Harb2018WhenWI}. Both of them have been employed in many single-agent applications. 

As depicted in \figurename~\ref{fig:intro}, various brain regions are responsible for various organs. We view different regions of the cerebral cortex as different high-level strategies. When an individual performs a complex action, coordination between high-level strategies is required. Similarly, we see the peripheral nervous system distributed in various organs as different low-level strategies. They also need to coordinate with each other (we ignore that coordination is mainly done through the spinal cord). Besides, high-level strategies need to guide their respective low-level strategies. Therefore, when viewing the human body as a multi-agent system, we find that inter-level and intra-level coordination are crucial for solving fully cooperative tasks. 

Inspired by the human nervous system, we propose a new framework for multi-agent cooperation problems, \textbf{H}ier\textbf{A}rchical \textbf{V}alue d\textbf{E}compositio\textbf{N} (HAVEN), a hierarchically structured value decomposition method. HAVEN develops a two-level QMIX-style strategy and uses the advantage function of the high-level policy as part of the low-level reward. In this way, the coordination of inter-level and inter-agent policies is guaranteed. There is also no need to pre-train the low-level policies. Simultaneously, because the action space of the high-level policies is preset while maintaining the generality, domain knowledge is not required for the training process of the entire framework. Besides, it should be noted that HAVEN can be extended to any value decomposition variant. In summary, HAVEN is an end-to-end and knowledge-free framework. 

Our contributions include two aspects: 
\begin{itemize}
    \item We present the HAVEN framework with a dual coordination mechanism of inter-level and inter-agent to solve the decentralized partially observable Markov decision process (Dec-POMDP) problems.
    \item Empirical evaluations in two testbeds, including StarCraft II~\cite{Samvelyan2019TheSM} and Google Research Football~\cite{Kurach2020GoogleRF}, also demonstrate that our method outperforms previous algorithms by a substantial margin.
\end{itemize}


\section{Preliminaries}

\subsection{Hierarchical Reinforcement Learning}

Hierarchical reinforcement learning is a structured framework intended to tackle complex problems by learning to make decisions over different levels of temporal abstraction. Since most of the related work is two hierarchy levels, we focus on the two-level structures. We call the whole hierarchical system the joint policy $\pi^{joint}$, composed of the high-level policy $\pi^h$ and low-level policy $\pi^l$. In the option-based hierarchical methods~\cite{Bacon2017TheOA, Harb2018WhenWI}, the action space of the high-level policy $\pi^h$ is discrete, and a low-level policy $\pi^l$ will be selected from a finite set of ones. For subgoal generation~\cite{Vezhnevets2017FeUdalNF, Nachum2018DataEfficientHR, Li2021LearningSR}, the output space of the high-level policy $\pi^h$ is often continuous. We need to calculate the intrinsic reward to guide the low-level policy $\pi^l$ to make decisions based on the goals generated by $\pi^h$. High-level strategies and low-level strategies often operate at two different time scales. One straightforward form~\cite{Zhang2021HierarchicalRL} is that $\pi^h$ runs every $k$ step to determine the low-level policies or subgoals in the next $k$ steps. Another more complicated approach~\cite{Rafati2019UnsupervisedMF} judges whether the subgoal is reached. If $\pi^l$ achieves the subgoal, $\pi^h$ makes a new decision and selects the next subgoal. Besides, we can set the termination function~\cite{Bacon2017TheOA, Harb2018WhenWI} to control whether or not $\pi^h$ makes a new decision.

\subsection{Value Decomposition Methods in Dec-POMDPs}

In this paper, we consider a fully cooperative multi-agent task that can be modelled by a Dec-POMDP~\cite{Oliehoek2016ACI}, which can be represented by the tuple $G=\langle S,U, A, P, r, Z, O, n, \gamma \rangle$. At each time step, each agent $a\in A:=\{1, \dots,n\}$ selects the corresponding action $u^a \in U$ with only having access to the local observation $z^a\in Z$ obtained by $O(s,a):S \times A \to Z$, where $s \in S$ is the real state of the environment. The joint action of all agents is defined as $\boldsymbol{u}\in \boldsymbol{U}$. The environmental dynamics, also known as the state transition function, is written as $P(s^\prime \mid s, \boldsymbol{u}):S\times \boldsymbol{U} \times S \to [0,1]$ . In Dec-POMDPs, all agents share a reward function: $r(s, \boldsymbol{u}): S\times \boldsymbol{U}\to \mathbb{R}$ . $\gamma \in [0, 1)$ is the discount factor. The goal of the multi-agent reinforcement learning problem in the Dec-POMDP is to maximize the discounted return $\sum_j^\infty\gamma^j r_{t+j}$.

An essential concept for multi-agent value decomposition methods is decomposability. Specifically, the overall and individual interests in the multi-agent system are consistent. This assumption can be formulated as Individual-Global-Max (IGM)~\cite{Son2019QTRANLT}, which assumes that the optimality of each agent $\arg \max _{u^a} Q_{a}\left(\tau^{a}, u^{a}\right)$ is consistent with the optimality of all agents $\arg \max _{u^a} Q_{tot}(\boldsymbol{\tau}, \boldsymbol{u})$. The equation that describes IGM is as follows:
\begin{equation*}
    \arg \max _{u^a} Q_{tot}(\boldsymbol{\tau}, \boldsymbol{u})=\arg \max _{u^a} Q_{a}\left(\tau^{a}, u^{a}\right), \quad \forall a \in A,
\end{equation*}
where $\boldsymbol{\tau} \in T^n$ represents the joint action-observation histories of all agents, $Q_{tot}$ is the global action-value function, and $Q_a$ is the individual one. Many variants of value decomposition have been developed and HAVEN can be applied to these methods.


\section{Method}

This section introduces the proposed novel multi-agent hierarchical reinforcement learning framework HAVEN. We first describe the entire process of HAVEN for interacting with the environment and then elaborate on its structure and implementation. Finally, we give the loss functions.

\begin{figure}[b]
    \centering
    \includegraphics[width=3.3 in]{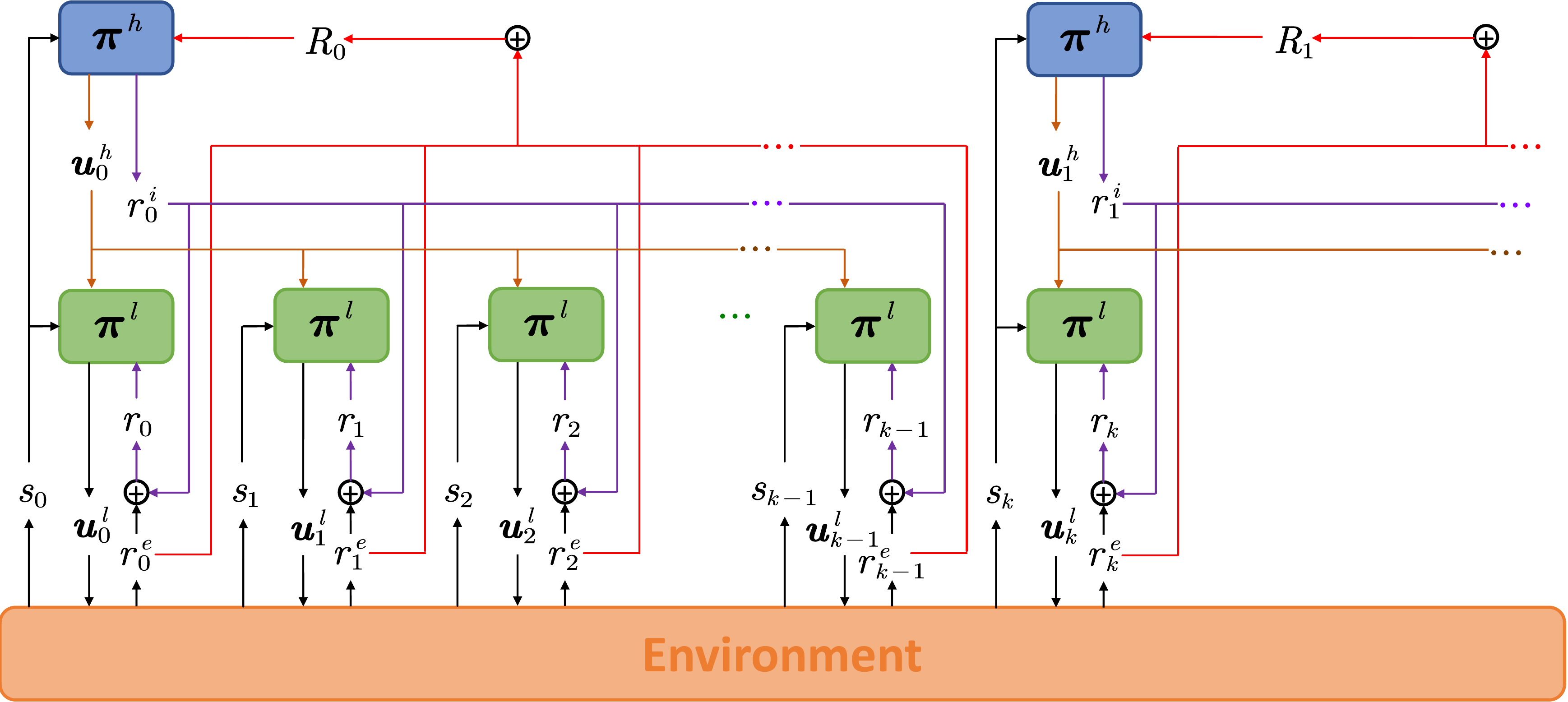}
    \caption{The workflow of HAVEN over an episode. The purple lines and the red lines represent the calculation processes of the reward function of $\boldsymbol{\pi}^l$ and $\boldsymbol{\pi}^h$, respectively.}
    \label{fig:Workflow}
\end{figure}

\subsection{The HAVEN Work Process}

In multi-agent systems, each agent $a$ has a high-level policy $\pi^{h,a}$ and a low-level policy $\pi^{l,a}$. And the corresponding action spaces are the macro action space $u^{h,a}\in U^h$ and the primitive action space $u^{l,a} \in U^l$. We define the macro action space $U^h$ as $N$ one-hot variables in this paper so that the output space of $\pi^{h,a}$ is discrete. $\boldsymbol{\pi}^h=\{\pi^{h,1},\dots,\pi^{h,n}\}$ represents the high-level joint policy of all agents, and $\boldsymbol{\pi}^l=\{\pi^{l,1},\dots,\pi^{l,n}\}$ denotes the low-level one.

HAVEN uses a two-timescale framework, faster for the low-level policy, and slower by a factor of $k$ for the high-level policy. So we define $T$ and $t$ as the time scales of the high-level policy and the low-level one, respectively. We carry out $\boldsymbol{\pi}^h$ every $k$ steps at the slow time scale. After $\boldsymbol{\pi}^h$ selects the joint macro action $\boldsymbol{u}^h$, $\boldsymbol{\pi}^l$ will select the joint primitive action $\boldsymbol{u}^l$ depending on the local observation $\boldsymbol{z}$ for $k$ steps. In Dec-POMDPs, all agents share a reward function given by environments and we denote it as the external reward $r^e$ of $\boldsymbol{\pi}^l$. We also set the high-level reward function to be shared, defined as $R_T=\sum_{i=0}^{k-1}r_{T\cdot k+i}^e$. We denote the replay buffers of the both level policies as $\mathcal{D}^l$ and $\mathcal{D}^h$ respectively, and the stored trajectories correspond to $\langle s_t,\boldsymbol{z}_t, \boldsymbol{u}^h_{\lfloor t/k\rfloor},\boldsymbol{u}^l_t, r^e_t\rangle$ and $\langle s_T,\boldsymbol{z}_T,\boldsymbol{u}^h_T, R_T \rangle$.

For the sake of the concurrent optimization of policies at both levels, we adopt the advantage function of $\boldsymbol{\pi}^h$ as the intrinsic reward of $\boldsymbol{\pi}^l$. An intuitive interpretation of the intrinsic reward is that the high-level advantage function can give low-level policies the temporal abstraction of next $k$ steps and guide them to learn skills. When $\boldsymbol{\pi}^h$ performs the joint action $\boldsymbol{u}^h_T$ in state $s_T$, we set the advantage function for $\boldsymbol{u}^h_T$ as $A_h(s_T,\boldsymbol{u}^h_T)$. Then for $\boldsymbol{\pi}^l$, the advantage function $A_h$ is evenly divided among $k$ steps to get the intrinsic reward of each low-level step, which can be expressed as:
\begin{equation}
    r^i_t=\frac{A_h(s_T,\boldsymbol{u}_T^h)}{k},\qquad T\cdot k\leq t<(T+1)\cdot k.
    \label{eq:intrisic_reward}
\end{equation}
$r^i$ links the strategies of different levels together. $R$ and $r^e$ act as joint reward functions between all agents. They respectively represent the coordination of the inter-level and inter-agent policies. Furthermore, according to Eq.~\eqref{eq:intrisic_reward}, the advantage-based intrinsic reward $r^i$ does not change in a $k$-step time interval. Equal rewards over a period of time can cause lower-level policies to suffer from temporal credit assignment problems. Agents need to know the actual feedback of the chosen action at each low-level step. Therefore, we get the linear combination of the external reward and the intrinsic reward simply to obtain the reward function $r=r^e+r^i$ of the low-level joint policy $\boldsymbol{\pi}^l$. The whole workflow of HAVEN is shown in Figure~\ref{fig:Workflow}. It should be noted that under the off-policy setting, the intrinsic reward is calculated during training, which means that $r^{i}$ is recalculated every time after sampling previous transitions from the replay buffer. So $r^{i}$ corresponding to each trajectory is not fixed. The calculation methodology of the intrinsic reward represented by the purple line in Figure~\ref{fig:Workflow} is only for the convenience of illustration.

\subsection{The HAVEN Framework}

To better address the credit assignment problem in Dec-POMDPs, policies at both levels in HAVEN are QMIX-style architectures, including a shared Agent Net and a Mixing Net. For the high-level policy, at the $T$-th high-level step, each agent $a$ chooses the macro action $u^{h,a}_T=\varepsilon\text{-greedy}\left(Q^h_a(\tau^{h,a}_T,u )\right)$ on the condition of the local observation $z^a_{T}$ and the previous macro action $u^{h,a}_{T-1}$. After all agents select the macro action through the Macro Agent Net composed of DRQN~\cite{Hausknecht2015DeepRQ}, the corresponding individual macro action-values will be fused by the Macro Mixing Net to obtain the global macro action-value $Q_{tot}^h(\boldsymbol{\tau}^h,\boldsymbol{u}^h)$. The specific structure of the Mixing Net is determined by the basic value decomposition method. Similarly, we obtain the joint primitive action $\boldsymbol{u}^l_t$ and the global low-level action-value function $Q_{tot}^l(\boldsymbol{\tau}^l, \boldsymbol{ u}^h, \boldsymbol{ u}^l)$ through the low-level value decomposition structure. However, the difference is that the input of the low-level Agent Net contains the macro action $u^{h,a}_T$ given by the Macro Agent Net.

\begin{figure}[tp]
    \centering
    \includegraphics[width=3.3 in]{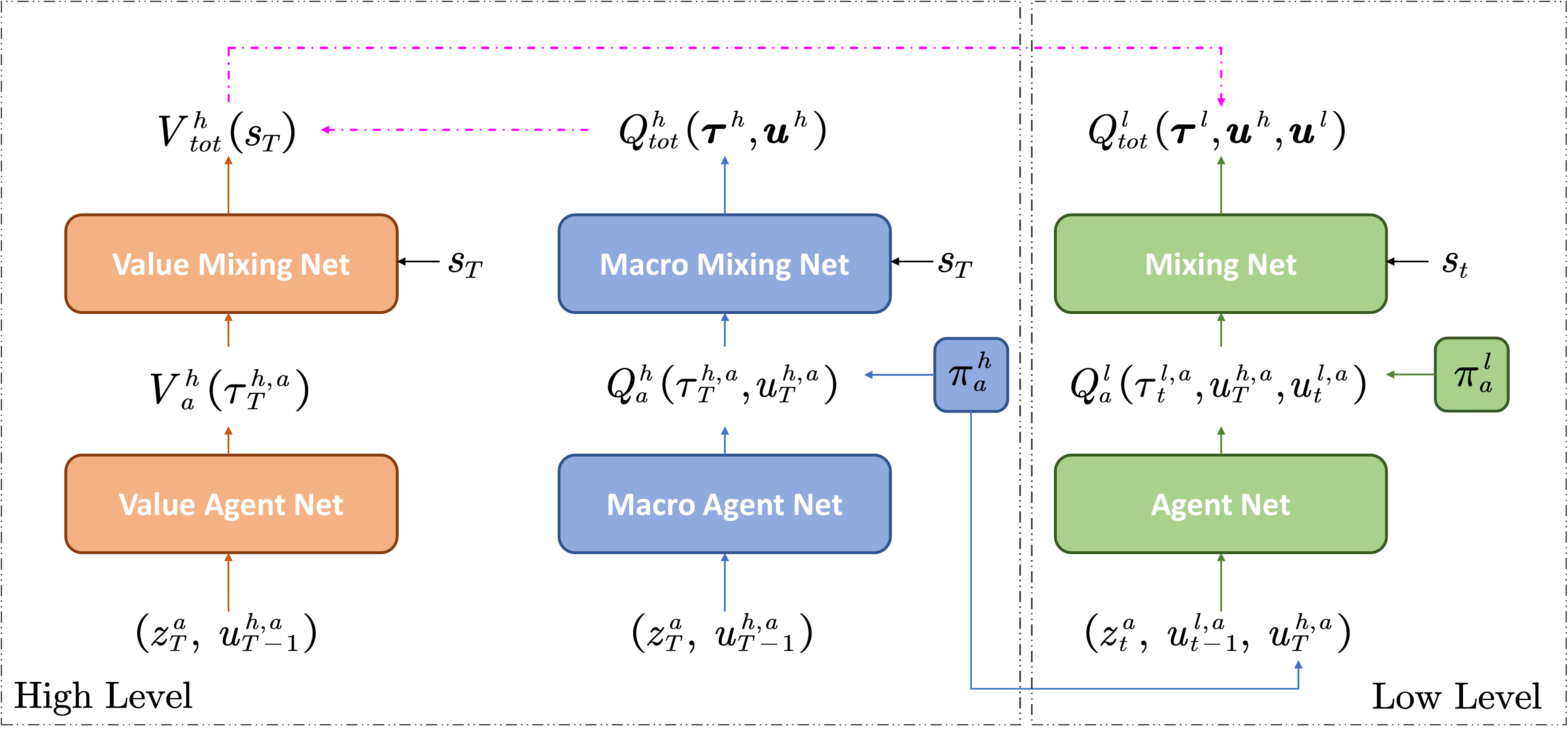}
    \caption{The overall HAVEN architecture. The left part is the high-level policy which includes $V^h(\cdot)$ and $Q^h(\cdot)$ two functions. And the right part is low-level policy which is the vanilla value decomposition architecture. The pink dashed arrows indicate that the update of one function is conditioned on another function.}
    \label{fig:framework}
\end{figure}

At the $T$-th high-level step, the advantage function can be defined as:
\begin{small}
\begin{equation}
A_h(s_T,\boldsymbol{u}^h_T)=\mathbb{E}_{s_{T+1}\sim(\boldsymbol{\pi^h},\boldsymbol{\pi^l})}\left[R_T+\gamma V^h(s_{T+1})-V^h(s_T)\right],
\label{eq:advantage}
\end{equation}
\end{small}
where $V^h(\cdot)$ represents the value function, which is usually estimated by the on-policy methods and gives the expected return if agents start in a certain state and always act according to the current policy $\boldsymbol{\pi}^{joint}$. Therefore, we need to add an additional neural network structure to estimate $V^h(\cdot)$. Enlightened by VDAC~\cite{Su2021ValueDecompositionMA}, we calculate the local state value $V_a^h(\tau_T^{h,a})$ of all agents and then feed them into the Value Mixing Net to finally get the global state-value function $V^h_{tot}(s_T)$. The additional neural network to approximate $V^h(\cdot)$ is indispensable.
We tried the advantage function $A(s, \boldsymbol{u}) = Q_{tot}(s, \boldsymbol{u}) - \max Q_{tot}(s,\cdot)$, but it didn't work well. The possible reason is that $Q^h_{tot}(s, \boldsymbol{u}^h)$ with respect to the non-optimal action $\boldsymbol{u}^h$ is inaccurately estimated. The state-value function $V^h(s)$ does not depend on the action and therefore is trained using more experiences than a action-value function that is only updated if a specific action is selected.
The overall framework of HAVEN is depicted in Figure~\ref{fig:framework}. Note that although two sets of neural networks have been added, the parameters of the entire framework did not increase linearly with the number of agents due to the parameter sharing mechanism. The number of parameters is similar to other multi-agent hierarchical reinforcement learning methods.

\subsection{Loss Functions}

HAVEN can realize the concurrent optimization of inter-level policies, similar to the monotonic improvement of joint policy in HAAR~\cite{Li2019HierarchicalRL}. Nevertheless, the most obvious difference is that HAVEN is an off-policy value-based method so that it can achieve higher sample efficiency. 
The objective function which is used by general on-policy methods for learning the state-value function in Eq.~\ref{eq:advantage} is given by the following equation:
\begin{equation}
    V^h(s_T)=(1-\alpha)V^h(s_T)+\alpha (R_T+\gamma V^h(s_{T+1})),
\label{eq:on_policy_v_update}
\end{equation}
where $\alpha$ is the learning rate. Eq.~\ref{eq:on_policy_v_update} obviously can not be directly applied to the off-policy reinforcement learning algorithm. The reason is that under the off-policy settings, the state value in Eq.~\ref{eq:on_policy_v_update} is estimated from the past policy and not from the target policy $\boldsymbol{\pi}^{joint}$ we need to optimize.
To address this issue, we have modified the update formula of the state-value function. The point is that the state-value estimates is obtained under the target policy. The state value of the $T+1$ step can be represented by the action value $\max_{\boldsymbol{u}^h} Q^h_{tot}(\cdot)$ under the target policy.
Enlightened by the loss function in QVMAX~\cite{Wiering2009TheQF}, we provide the off-policy objective function by using the $\max$ operator:
\begin{small}
\begin{equation}
    V^h(s_T)=(1-\alpha)V^h(s_T)+\alpha (R_T+\gamma \max_{\boldsymbol{u}^h_{T+1}} Q_{tot}^h(\boldsymbol{\tau}^h_{T+1},\boldsymbol{u}^h_{T+1})).
\label{eq:off_policy_v_update}
\end{equation}
\end{small}
$R_T$ in Eq.~\ref{eq:off_policy_v_update} is obtained under the past behavior policy instead of the target policy $\boldsymbol{\pi}^{joint}$. So the resulting algorithm uses a one-step trajectory of the behavior policy, which means that the state value function $V^h(\cdot)$ under the joint policy $\boldsymbol{\pi}^{joint}$ calculated by Eq.~\ref{eq:off_policy_v_update} is still different from the true state value. However, compared with Eq.~\ref{eq:on_policy_v_update} in which $R_T$ and $V^h(\cdot)$ are all estimated from the past policy, our proposed modified objective function is closer to the true value under the off-policy settings.

We take the initial state value $\eta$ as the optimization goal, which means we need to find a suitable joint policy $\boldsymbol{\pi}^{joint}$ to maximize it. The equation that describes $\eta(\boldsymbol{\pi}^{joint})$ is as follows:
\begin{small}
\begin{equation}
\begin{split}
\eta(\boldsymbol{\pi}^{joint}) &=\mathbb{E}_{s_{0}^{h}}\left[V^h\left(s_{0}^{h}\right)\right] \\
&=\mathbb{E}_{s_{0}^{h}, \boldsymbol{u}_{0}^{h}, \ldots \sim \boldsymbol{\pi}^{joint}}\left[\sum_{T} \gamma_{h}^{T} R\left(s_{T}^{h}\right)\right].
\end{split}
\label{eq:pi_target}
\end{equation}
\end{small}
We can easily get that, in the case of fixed low-level policy $\boldsymbol{\pi}^l$, optimizing high-level policy $\boldsymbol{\pi}^h$ leads to improvement in the joint policy $\boldsymbol{\pi}^{joint}$. For the optimization of the low-level policy, we need to justify it theoretically. We assume that the high-level policy is fixed when we optimize the low-level policy, and use $\tilde{\boldsymbol{\pi}}^{joint}$ and $\tilde{\boldsymbol{\pi}}^{l}$ to represent the updated joint policy and the updated low-level policy. We can obtain the optimization target of $\tilde{\boldsymbol{\pi}}^{joint}$ and $\tilde{\boldsymbol{\pi}}^{l}$:
\begin{small}
\begin{align}
&\eta(\tilde{\boldsymbol{\pi}}^{joint})=\eta(\boldsymbol{\pi}^{joint})\nonumber\\&+\mathbb{E}_{\left(s_{T}^{h}, \boldsymbol{u}_{T}^{h}\right) \sim \tilde{\boldsymbol{\pi}}^{joint}}\left[\sum_{T} \gamma_{h}^{T} A_{h}\left(s_{T}^{h}, \boldsymbol{u}_{T}^{h}\right)\right],\label{eq:tilde_pi_target}\\
&\eta\left(\tilde{\boldsymbol{\pi}}^{l}\right)\approx\eta(\boldsymbol{\pi}^{joint})\nonumber\\&+\left[1+\frac{1-\gamma_{l}^{k}}{k(1-\gamma_{l})}\right]\mathbb{E}_{\tau_{h} \sim\left(\tilde{\boldsymbol{\pi}}_{l}, \boldsymbol{\pi}_{h}\right)}\left[\sum_{T} \gamma_{h}^{T} A_{h}\left(s_{T}^{h}, \boldsymbol{u}_{T}^{h}\right)\right].\label{eq:tilde_pi_l_target}
\end{align}
\end{small}
The proof of Eq.~\eqref{eq:tilde_pi_l_target} can be found in Appendix A. When the two-level discount factors $\gamma_l$ and $\gamma_h$ are close to 1 and $k$ is not large, the optimization goals of the joint policy and that of the low-level policy have both the term $\mathbb{E}_{\boldsymbol{\tau}_{h} \sim\left(\tilde{\boldsymbol{\pi}}_{l}, \boldsymbol{\pi}_{h}\right)}\left[\sum_{T} \gamma_{h}^{T} A_{h}\left(s_{T}^{h}, \boldsymbol{u}_{T}^{h}\right)\right]$. Meantime, since the updated policy has nothing to do with the original joint policy $\boldsymbol{\pi}^{joint}$, the optimization goals of the two can be further simplified:
\begin{small}
\begin{equation*}
\begin{aligned}
&\max _{\tilde{\pi}^{joint}} \eta(\tilde{\pi}^{joint})=\max_{\tilde{\pi}^{joint}} \mathbb{E}_{\left(s_{t}^{h}, \boldsymbol{u}_{t}^{h}\right) \sim \tilde{\pi}}\left[\sum_{T} \gamma_{h}^{T} A_{h}\left(s_{T}^{h}, \boldsymbol{u}_{T}^{h}\right)\right],\\
&\max_{\tilde{\pi}^l}\eta({\tilde{\pi}^l})=\\&\max_{\tilde{\pi}^l}\left[1+\frac{1-\gamma_{l}^{k}}{k(1-\gamma_{l})}\right]\mathbb{E}_{\tau_{h} \sim\left(\tilde{\pi}^{l}, \pi^{h}\right)}\left[\sum_{T} \gamma_{h}^{T} A_{h}\left(s_{T}^{h}, \boldsymbol{u}_{T}^{h}\right)\right].
\end{aligned}
\end{equation*}
\end{small}
$1+\frac{1-\gamma_l^k}{k(1-\gamma_l)}$ is obviously a positive value, so when we maximize Eq.~\eqref{eq:tilde_pi_l_target}, $\eta(\tilde{\boldsymbol{\pi}}^{joint})$ in Eq.~\eqref{eq:tilde_pi_target} increases. To sum up, the joint policy $\boldsymbol{\pi}^{joint}$ is monotonically optimized when we monotonically optimize $\boldsymbol{\pi}^h$ and $\boldsymbol{\pi}^l$. So the above updating scheme with the intrinsic reward avoids instability of the concurrent optimization of inter-level policies.
In addition to the intrinsic reward $r^i$ that focuses on the coordination of inter-level policies, the reward function of $\boldsymbol{\pi}^l$ includes the external rewards $r^e$, which can improve cooperation between agents through the value decomposition mechanism of the low-level policy and alleviate temporal credit assignment problems mentioned above. 

From the above explanation, we can get the loss function of the three sets of neural networks: the high-level state-value network, the high-level action-value network, and the low-level action-value network. $\theta$, $\phi$, and $\psi$ represent their parameters, respectively. Thus, the following loss function is obtained:
\begin{small}
\begin{align}
    \mathcal{L}_V^h(\theta)=&\bigg(R_T+\gamma_h \max_{{\boldsymbol{u}^h_{T+1}}}Q_{tot}^h\left({\boldsymbol{\tau}^h_{T+1}},{\boldsymbol{u}^h_{T+1}}\mid \phi\right)\nonumber\\
    &-V^h\left(s_{T} \mid \theta\right)\bigg)^2,\\
    \mathcal{L}_Q^h(\phi)=&\bigg(R_T+\gamma_h\max_{{\boldsymbol{u}^h_{T+1}}} Q_{tot}^h\left({\boldsymbol{\tau}^h_{T+1}},{\boldsymbol{u}^h_{T+1}}\mid \phi^-\right)\nonumber\\
    &-Q^h_{tot}\left(\boldsymbol{\tau}^h_{T},\boldsymbol{u}^h_{T}\mid \phi\right)\bigg)^2,\\
    \mathcal{L}_{Q}^l(\psi)=&\bigg(r_t+\gamma_l \max_{{\boldsymbol{u}^l_{t+1}}}Q_{tot}^l\left({\boldsymbol{\tau}^l_{t+1}},{\boldsymbol{u}^h_{\lfloor (t+1)/k\rfloor}}, {\boldsymbol{u}^l_{t+1}} \mid \psi^-\right)\nonumber\\
    &-Q_{tot}^l\left(\boldsymbol{\tau}^l_{t},{\boldsymbol{u}^h_{\lfloor t/k\rfloor}}, {\boldsymbol{u}^l_{t}} \mid \psi\right)\bigg)^2,
\end{align}
\end{small}
where $\phi^-$ and $\psi^-$ refer to the parameters of the high-level and low-level action-value target network, respectively. It is worth mentioning that the optimization of the three networks is independent of each other. The implementation details and algorithmic description of HAVEN can be found in Appendix B.


\section{Related Work}

Several approaches to single-agent HRL have been proposed. One is the method based on options, which abstracts frequently reused sub-policies into actions of the high-level policy. This approach often causes options to degenerate into primitive actions. A somewhat different approach identifies a set of representations (usually the subset of the state space or the hidden variable space) that make for useful subgoals. The output space of the high-level policy is set to the subgoal space, and the low-level policy outputs the primitive actions depending on the subgoals output by the high-level policy. Although this approach is quite similar to human decision-making, it is frequently challenging to put into practice because of the size of the subgoal space. We can also speed up the reinforcement learning process by manually setting subgoals~\cite{Rafati2019LearningRI,Song2019PlayingFG} or intrinsic rewards~\cite{Vezhnevets2017FeUdalNF}, but this inevitably introduces the domain knowledge. In addition, some methods have been proposed to solve the instability caused by the simultaneous learning of policies at both levels. For example, HAAR~\cite{Li2019HierarchicalRL} calculates advantage-based auxiliary rewards, and CHER~\cite{Kreidieh2019InterLevelCI} collaboratively optimizes goal-assignment and goal-achievement policies from a multi-agent perspective. 

In recent years, hierarchical structures have gradually been used in multi-agent reinforcement learning. Feudal Multi-agent Hierarchies (FMH)~\cite{Ahilan2019FeudalMH} applies a feudal architecture to the multi-agent environments. However, its primary flaw is that it cannot be applied to a fully-cooperative setting, in which all agents optimize a shared reward function. To address the sparse and delayed reward problem in the cooperative multi-agent situation, the hierarchical deep multi-agent reinforcement learning methods with temporal abstraction~\cite{Tang2018HierarchicalDM} such as Hierarchical QMIX and Hierarchical Communication Network were proposed. Nevertheless, the significant limitation of these methods is that the high-level action space is set manually. Hierarchical learning with skill discovery (HSD)~\cite{Yang2020HierarchicalCM} makes the skills output by the macro policy more diversified through supervised learning, which helps agents learn useful skills. However, the parameter tuning process in HSD is tedious because of a large number of hyperparameters. RODE~\cite{Wang2021RODELR} explicitly divides the action space by clustering actions. Each action subspace corresponds to a kind of ``role". It is a novel idea except for the high cost of clustering. \looseness=-1

Our proposed HAVEN is \textbf{NOT} a method that simply replaces the intrinsic reward in HSD with a HAAR-like advantage-based objective. The high-level Q-function in HSD is a QMIX architecture but the low-level one is learned by independent Q-learning. Due to the belief that low-level policies would continue to experience credit assignment issues, we suggest bi-level QMIX-style structures for HAVEN. Furthermore, the low-level reward is defined as a combination of the team reward and the intrinsic reward in HSD, where the latter is a probability value. It is difficult for the skill discovery mechanism to work if the scales of the team reward and the intrinsic reward are not essentially equivalent. However, the scales of $r^e$ and $r^i$ in the low-level reward in HAVEN must be the same. In HSD, the intrinsic reward is calculated when interacting with the environment and stored directly in the replay buffer. While in HAVEN the intrinsic reward is calculated at training time, so it does not become obsolete as the policies change.

HAAR is an on-policy Actor-Critic algorithm. To improve the sample efficiency, we modify its loss function and successfully apply the advantage-based intrinsic reward to off-policy multi-agent value decomposition methods. HAAR needs to pre-train low-level skills while HAVEN does not. Simultaneously, the low-level reward in HAVEN is not just the high-level advantage function like HAAR, we also introduce external reward to alleviate the temporal credit assignment problem caused by equal intrinsic rewards in a $k$-step time interval. And we theoretically prove that the addition of external reward does not affect the monotonic improvement of the inter-level policies in Appendix A.


\section{Experiments}

In this section, we test our method on the StarCraft II micromanagement benchmark and the Google Research Football environment. Then by carrying out ablation studies, we show that each module that constitutes HAVEN is not redundant. We also investigate the influence of different hyperparameter settings. Finally, we visualize both level policies, which sheds further light on the role of the dual coordination mechanism. Details of the experimental setup can be found in Appendix C.

\begin{figure*}[ht]
    \centering
    \includegraphics[width=5.5 in]{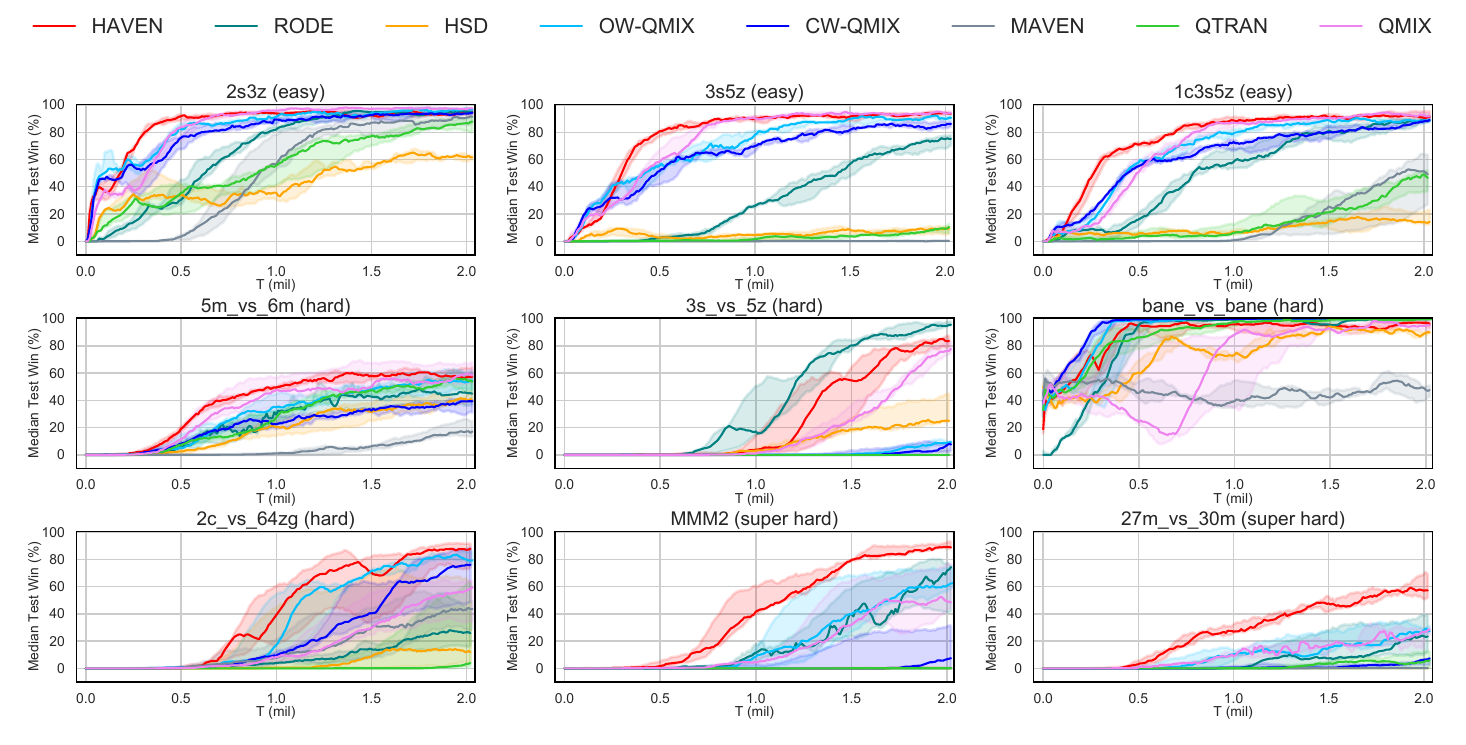}
    \caption{Performance comparison with baselines in different scenarios.}
    \label{fig:results}
\end{figure*}

\subsection{Performance on StarCraft II}

We first evaluate the performance of HAVEN in the SMAC testbed and compare it with other popular baselines. SMAC is a multi-agent reinforcement learning environment based on the real-time strategy game StarCraft II. There are numerous Dec-POMDP micromanagement tasks in SMAC. The version of StarCraft II is SC2.4.6.2.69232 which is the same as some literature~\cite{Rashid2018QMIXMV, Rashid2020WeightedQE}, not the easier SC2.4.10. Performance is not always comparable between versions. To verify the validity of our method, we choose the most common method QMIX as the basic algorithm of HAVEN. Of course, HAVEN can also be built based on other value decomposition algorithms.

\begin{figure*}[h]
    \centering
    \includegraphics[width=6.0 in]{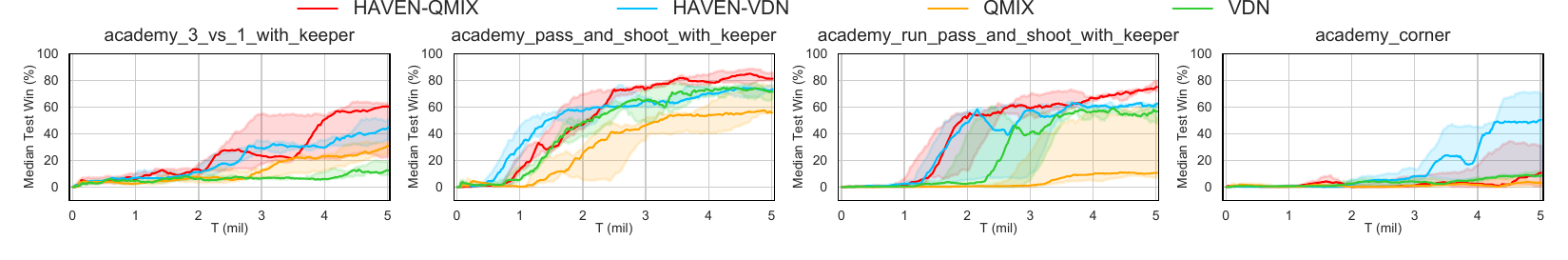}
    \caption{Comparison of our approach against baseline algorithms on Google Research Football.}
    \label{fig:grf_results}
\end{figure*}

The implementation of HAVEN and other benchmarks in our experiment is based on Pymarl~\cite{Samvelyan2019TheSM}. To make the empirical results more convincing, we compare HAVEN with state-of-the-art value decomposition approaches, including QMIX, QTRAN~\cite{Son2019QTRANLT}, Weighted QMIX~\cite{Rashid2020WeightedQE}, MAVEN~\cite{Mahajan2019MAVENMV}, and following hierarchical multi-agent reinforcement learning methods: 
\begin{itemize}
    \item \textbf{RODE}~\cite{Wang2021RODELR}\quad which decomposes joint action spaces based on the high-level role selector. RODE did some changes to the SMAC environment but did not carry over these changes to the baselines in the original paper. We test it under the original environment settings and show its real performance.
    \item \textbf{HSD}~\cite{Yang2020HierarchicalCM}\quad which is a hierarchical multi-agent method with unsupervised skill discovery for strategic teamwork.
\end{itemize}
The hyperparameters of the basic algorithm in HAVEN and those of other baseline algorithms are consistent with the original work. For the high-level time scale $k$ and the number of macro actions $N$, we set them to 3 and 8, respectively.

Figure~\ref{fig:results} shows the performance comparison between HAVEN based on QMIX and other baselines in different scenarios. The solid lines represent the median win rates, and the 25-75\% percentiles are shaded. The performance of our method is significantly better than its basic algorithm QMIX and many other baselines. The role-based learning of RODE does not bring much performance improvement in our fair comparison. HSD also performs poorly and we guess it is because in the original paper HSD was only evaluated in a fully observable game. So the skill discovery mechanism in HSD may not work in a partially observable environment, even though we used DRQN. The superiority of HAVEN is more obvious in hard scenario \emph{2c\_vs\_64zg}, and super hard scenarios \emph{MMM2} and \emph{27m\_vs\_30m}. In addition, HAVEN can still achieve high sample efficiency in all easy scenarios, which is difficult for some other complex value decomposition methods. The dual coordination mechanism in HAVEN significantly improves the sample efficiency, and we will analyze it further in the visualization.

\subsection{Performance on Google Research Football}

The Google Research Football environment is a reinforcement learning experimental platform focused on training agents to play football. We study the effectiveness and generalization of HAVEN framework in the Football Academy, which has many mini-scenarios. We tried two different basic algorithms, VDN and QMIX. These methods applied with HAVEN are denoted as HAVEN-VDN and HAVEN-QMIX, respectively. We evaluate them and the vanilla algorithms on some official maps of Google Research Football. All other experimental settings are the same as those on SMAC.

Figure~\ref{fig:grf_results} shows the learning curve on four different maps. In all scenarios, both VDN and QMIX, their performance is worse than HAVEN-VDN and HAVEN-QMIX. The overall experiments show that HAVEN can be applied to different Dec-POMDP domains and extended to different value decomposition algorithms.

\begin{figure*}[t]
\centering
\subfigure{
    \includegraphics[width=2.012409 in]{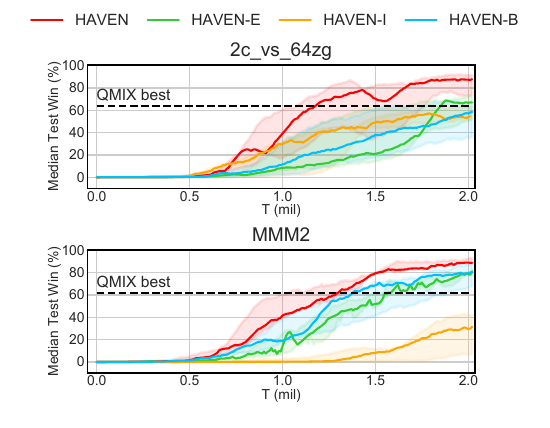}
}
\hspace{0 in}
\subfigure{
    \includegraphics[width=1.756287 in]{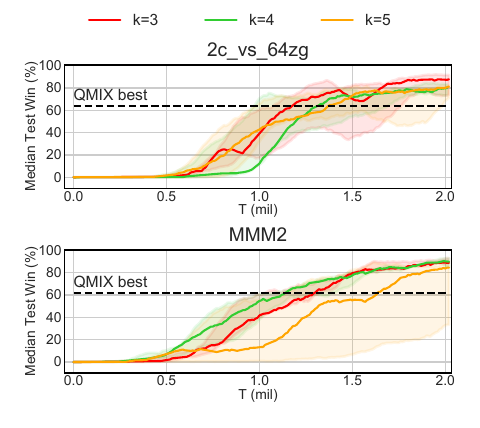}
}
\hspace{0.16 in}
\subfigure{
    \includegraphics[width=1.756287 in]{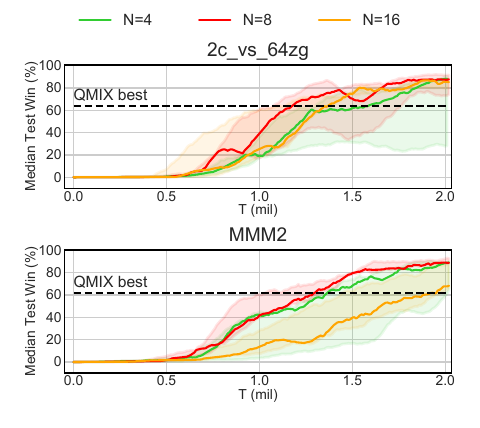}
}
\caption{\textbf{Left}: Win rates for HAVEN and ablations. \textbf{Middle}: Influence of the $k$ for HAVEN. \textbf{Right}: HAVEN with different $N$.}
\label{fig:extra_experiment}
\end{figure*}

\subsection{Ablation Studies}

We also carry out some ablation studies and discuss the influence of different values of $k$ and $N$. The ablation experiments include setting the low-level reward function to (1) only the intrinsic reward $r^i $ or (2) only the external reward $r^e$, and (3) using the general bootstrap update formula which is described by Eq.~\eqref{eq:on_policy_v_update} as the update formula of $V^h(\cdot)$ instead of the off-policy update mode in Eq.~\eqref{eq:off_policy_v_update}. We propose three ablations in which one of the above three components is different from the original HAVEN while the other parts remain unchanged. The above three alterations of HAVEN are called HAVEN-I, HAVEN-E, and HAVEN-B. Recall that the high-level policy is executed every $k$ steps and the number of macro actions is $N$. For the $k$ and $N$, we choose different values of the two to explore how they influence the performance. \looseness=-1

Ablation experiments are carried out on two typical scenarios in SMAC, \emph{2c\_vs\_64zg} and \emph{MMM2}. From Figure~\ref{fig:extra_experiment}, it can be seen that no matter which ablations, its performance is significantly worse than that of vanilla HAVEN. Especially in scenario \emph{MMM2}, the performance of HAVEN-I that only contains the intrinsic reward for the low-level policy is far worse than other ablations and the original framework, which means that the understanding the relationship between primitive actions and rewards greatly influences learning. So the introduction of external rewards in the low-level reward function can alleviate the above temporal credit assignment problem. Meanwhile, HAVEN-E does not perform well in \emph{2c\_vs\_64zg} because of the lack of the long-term intrinsic reward calculated from high-level value functions, which can guide the learning of low-level skills. Through the above analysis, both the intrinsic reward and the external reward in the low-level reward are indispensable. Comparing the performance of HAVEN-B and original HAVEN, we can also conclude that the inaccurate state-value function estimated in an on-policy manner is harmful to learning. As mentioned above, these three components all contribute to HAVEN.\looseness=-1

We show the influence of various hyperparameter settings on the performance of HAVEN in Figure~\ref{fig:extra_experiment}. First, we discuss how the $k$ influences the performance. The results show a trend that HAVEN performs worse as the $k$ increases, and this phenomenon is more pronounced in \emph{MMM2}. The experimental results match the assumption made in the previous section: $k$ cannot be very large. Regarding the setting of the number of macro actions $N$, we found that $N$ should not be too large either. We hypothesize that this is because a large $N$ value will enlarge the high-level action space and the low-level state space.

\subsection{Visualization}

\begin{figure}[t]
    \centering
    \includegraphics[width=2.7 in]{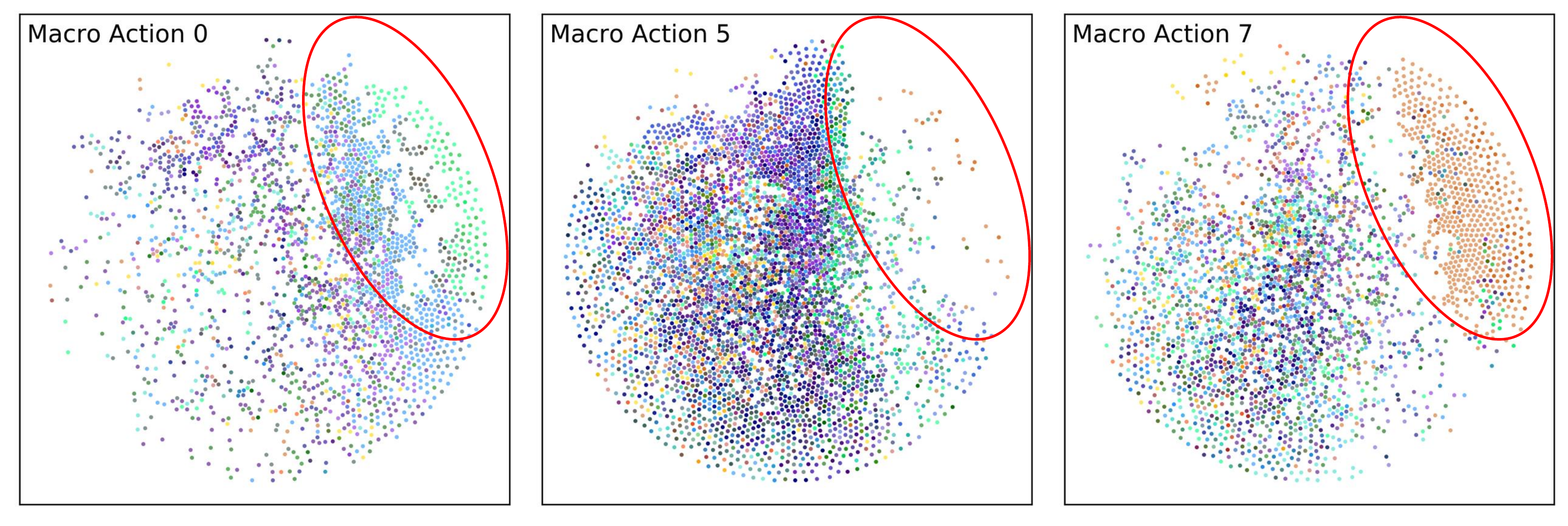}
    \caption{The 2D t-SNE embedding of states of the corresponding macro action. The colors of the points correspond to primitive actions.}
    \label{fig:action_pairs}
\end{figure}

According to previous work, it is straightforward to know that the intra-level policies implemented by the value decomposition structure are coordinated with each other and we visualize them in Appendix D. Therefore we focus on the visualization of inter-level policies. Figure~\ref{fig:action_pairs} shows the 2D t-SNE~\cite{Maaten2008VisualizingDU} embedding of states of the corresponding macro action selected by agents. The density of points in a particular area reflects agents' preference for the macro action in the corresponding state. Each point is colored according to primitive actions. We set the movement-related primitive actions as warm colors and attack-related actions as cool colors. It can be clearly seen that the regions in the red circles corresponding to the three different macro actions present three completely different situations. When in the states indicated by the red circle area, agents choosing Macro Action 0 are more inclined to attack, and those choosing Macro Action 7 are more likely to move. Agents rarely choose Macro Action 5 in these states. The coordination mechanism in HAVEN for inter-level policies guides lower-level policies to learn different skills for different macro actions.

\section{Conclusion}

This paper proposes a novel hierarchical off-policy value decomposition framework HAVEN, which is simple yet effective and can be applied to any value decomposition variant. The dual coordination mechanism for the simultaneous learning of inter-level and inter-agent policies also provides a solid theoretical foundation for the excellent performance of HAVEN. HAVEN does not need to set high-level action spaces manually and carry out pre-training. The experimental results show that HAVEN is robust to both easy and hard scenarios. We believe that our proposed HAVEN framework provides a general and efficient approach for multi-agent hierarchical reinforcement learning.

In our future research, we intend to concentrate on how to select the number of macro actions and whether the trained low-level policies can be transferred to other tasks. Further study of the issues would be of interest.


\bibliography{main.bbl}

\begin{thebibliography}{40}
\providecommand{\natexlab}[1]{#1}

\bibitem[{Ahilan and Dayan(2019)}]{Ahilan2019FeudalMH}
Ahilan, S.; and Dayan, P. 2019.
\newblock Feudal Multi-Agent Hierarchies for Cooperative Reinforcement
  Learning.
\newblock \emph{ArXiv}, abs/1901.08492.

\bibitem[{Bacon, Harb, and Precup(2017)}]{Bacon2017TheOA}
Bacon, P.; Harb, J.; and Precup, D. 2017.
\newblock The Option-Critic Architecture.
\newblock In Singh, S.~P.; and Markovitch, S., eds., \emph{Proceedings of the
  Thirty-First {AAAI} Conference on Artificial Intelligence, February 4-9,
  2017, San Francisco, California, {USA}}, 1726--1734. {AAAI} Press.

\bibitem[{Choi et~al.(2018)Choi, Ha, Hwang, Kim, Ha, and
  Yoon}]{Choi2018ReinforcementLB}
Choi, S.; Ha, H.; Hwang, U.; Kim, C.; Ha, J.-W.; and Yoon, S. 2018.
\newblock Reinforcement Learning based Recommender System using Biclustering
  Technique.
\newblock \emph{ArXiv}, abs/1801.05532.

\bibitem[{Dayan and Hinton(1992)}]{Dayan1992FeudalRL}
Dayan, P.; and Hinton, G.~E. 1992.
\newblock Feudal Reinforcement Learning.
\newblock In \emph{NIPS}.

\bibitem[{Dietterich(2000)}]{Dietterich2000HierarchicalRL}
Dietterich, T.~G. 2000.
\newblock Hierarchical Reinforcement Learning with the MAXQ Value Function
  Decomposition.
\newblock \emph{ArXiv}, cs.LG/9905014.

\bibitem[{Foerster et~al.(2016)Foerster, Assael, de~Freitas, and
  Whiteson}]{Foerster2016LearningTC}
Foerster, J.~N.; Assael, Y.~M.; de~Freitas, N.; and Whiteson, S. 2016.
\newblock Learning to Communicate with Deep Multi-Agent Reinforcement Learning.
\newblock In Lee, D.~D.; Sugiyama, M.; von Luxburg, U.; Guyon, I.; and Garnett,
  R., eds., \emph{Advances in Neural Information Processing Systems 29: Annual
  Conference on Neural Information Processing Systems 2016, December 5-10,
  2016, Barcelona, Spain}, 2137--2145.

\bibitem[{Foerster et~al.(2018)Foerster, Farquhar, Afouras, Nardelli, and
  Whiteson}]{Foerster2018CounterfactualMP}
Foerster, J.~N.; Farquhar, G.; Afouras, T.; Nardelli, N.; and Whiteson, S.
  2018.
\newblock Counterfactual Multi-Agent Policy Gradients.
\newblock In McIlraith, S.~A.; and Weinberger, K.~Q., eds., \emph{Proceedings
  of the Thirty-Second {AAAI} Conference on Artificial Intelligence, (AAAI-18),
  the 30th innovative Applications of Artificial Intelligence (IAAI-18), and
  the 8th {AAAI} Symposium on Educational Advances in Artificial Intelligence
  (EAAI-18), New Orleans, Louisiana, USA, February 2-7, 2018}, 2974--2982.
  {AAAI} Press.

\bibitem[{Harb et~al.(2018)Harb, Bacon, Klissarov, and Precup}]{Harb2018WhenWI}
Harb, J.; Bacon, P.; Klissarov, M.; and Precup, D. 2018.
\newblock When Waiting Is Not an Option: Learning Options With a Deliberation
  Cost.
\newblock In McIlraith, S.~A.; and Weinberger, K.~Q., eds., \emph{Proceedings
  of the Thirty-Second {AAAI} Conference on Artificial Intelligence, (AAAI-18),
  the 30th innovative Applications of Artificial Intelligence (IAAI-18), and
  the 8th {AAAI} Symposium on Educational Advances in Artificial Intelligence
  (EAAI-18), New Orleans, Louisiana, USA, February 2-7, 2018}, 3165--3172.
  {AAAI} Press.

\bibitem[{Hausknecht and Stone(2015)}]{Hausknecht2015DeepRQ}
Hausknecht, M.; and Stone, P. 2015.
\newblock Deep Recurrent Q-Learning for Partially Observable MDPs.
\newblock In \emph{AAAI Fall Symposia}.

\bibitem[{Iqbal and Sha(2019)}]{Iqbal2019ActorAttentionCriticFM}
Iqbal, S.; and Sha, F. 2019.
\newblock Actor-Attention-Critic for Multi-Agent Reinforcement Learning.
\newblock In Chaudhuri, K.; and Salakhutdinov, R., eds., \emph{Proceedings of
  the 36th International Conference on Machine Learning, {ICML} 2019, 9-15 June
  2019, Long Beach, California, {USA}}, volume~97 of \emph{Proceedings of
  Machine Learning Research}, 2961--2970. {PMLR}.

\bibitem[{Kreidieh et~al.(2019)Kreidieh, Parajuli, Lichtl{\'e}, You, Nasr, and
  Bayen}]{Kreidieh2019InterLevelCI}
Kreidieh, A.~R.; Parajuli, S.; Lichtl{\'e}, N.; You, Y.; Nasr, R.; and Bayen,
  A. 2019.
\newblock Inter-Level Cooperation in Hierarchical Reinforcement Learning.
\newblock \emph{ArXiv}, abs/1912.02368.

\bibitem[{Kurach et~al.(2020)Kurach, Raichuk, Stanczyk, Zajac, Bachem,
  Espeholt, Riquelme, Vincent, Michalski, Bousquet, and
  Gelly}]{Kurach2020GoogleRF}
Kurach, K.; Raichuk, A.; Stanczyk, P.; Zajac, M.; Bachem, O.; Espeholt, L.;
  Riquelme, C.; Vincent, D.; Michalski, M.; Bousquet, O.; and Gelly, S. 2020.
\newblock Google Research Football: A Novel Reinforcement Learning Environment.
\newblock In \emph{AAAI}.

\bibitem[{Kuyer et~al.(2008)Kuyer, Whiteson, Bakker, and
  Vlassis}]{Kuyer2008MultiagentRL}
Kuyer, L.; Whiteson, S.; Bakker, B.; and Vlassis, N. 2008.
\newblock Multiagent Reinforcement Learning for Urban Traffic Control Using
  Coordination Graphs.
\newblock In \emph{ECML/PKDD}.

\bibitem[{Li et~al.(2019)Li, Wang, Tang, and Zhang}]{Li2019HierarchicalRL}
Li, S.; Wang, R.; Tang, M.; and Zhang, C. 2019.
\newblock Hierarchical Reinforcement Learning with Advantage-Based Auxiliary
  Rewards.
\newblock In Wallach, H.~M.; Larochelle, H.; Beygelzimer, A.;
  d'Alch{\'{e}}{-}Buc, F.; Fox, E.~B.; and Garnett, R., eds., \emph{Advances in
  Neural Information Processing Systems 32: Annual Conference on Neural
  Information Processing Systems 2019, NeurIPS 2019, December 8-14, 2019,
  Vancouver, BC, Canada}, 1407--1417.

\bibitem[{Li et~al.(2021)Li, Zheng, Wang, and Zhang}]{Li2021LearningSR}
Li, S.; Zheng, L.; Wang, J.; and Zhang, C. 2021.
\newblock Learning Subgoal Representations with Slow Dynamics.
\newblock In \emph{ICLR}.

\bibitem[{Lowe et~al.(2017)Lowe, Wu, Tamar, Harb, Abbeel, and
  Mordatch}]{Lowe2017MultiAgentAF}
Lowe, R.; Wu, Y.; Tamar, A.; Harb, J.; Abbeel, P.; and Mordatch, I. 2017.
\newblock Multi-Agent Actor-Critic for Mixed Cooperative-Competitive
  Environments.
\newblock In Guyon, I.; von Luxburg, U.; Bengio, S.; Wallach, H.~M.; Fergus,
  R.; Vishwanathan, S. V.~N.; and Garnett, R., eds., \emph{Advances in Neural
  Information Processing Systems 30: Annual Conference on Neural Information
  Processing Systems 2017, December 4-9, 2017, Long Beach, CA, {USA}},
  6379--6390.

\bibitem[{Mahajan et~al.(2019)Mahajan, Rashid, Samvelyan, and
  Whiteson}]{Mahajan2019MAVENMV}
Mahajan, A.; Rashid, T.; Samvelyan, M.; and Whiteson, S. 2019.
\newblock {MAVEN:} Multi-Agent Variational Exploration.
\newblock In Wallach, H.~M.; Larochelle, H.; Beygelzimer, A.;
  d'Alch{\'{e}}{-}Buc, F.; Fox, E.~B.; and Garnett, R., eds., \emph{Advances in
  Neural Information Processing Systems 32: Annual Conference on Neural
  Information Processing Systems 2019, NeurIPS 2019, December 8-14, 2019,
  Vancouver, BC, Canada}, 7611--7622.

\bibitem[{Nachum et~al.(2018)Nachum, Gu, Lee, and
  Levine}]{Nachum2018DataEfficientHR}
Nachum, O.; Gu, S.; Lee, H.; and Levine, S. 2018.
\newblock Data-Efficient Hierarchical Reinforcement Learning.
\newblock In Bengio, S.; Wallach, H.~M.; Larochelle, H.; Grauman, K.;
  Cesa{-}Bianchi, N.; and Garnett, R., eds., \emph{Advances in Neural
  Information Processing Systems 31: Annual Conference on Neural Information
  Processing Systems 2018, NeurIPS 2018, December 3-8, 2018, Montr{\'{e}}al,
  Canada}, 3307--3317.

\bibitem[{Oliehoek and Amato(2016)}]{Oliehoek2016ACI}
Oliehoek, F.; and Amato, C. 2016.
\newblock A Concise Introduction to Decentralized POMDPs.
\newblock In \emph{SpringerBriefs in Intelligent Systems}.

\bibitem[{Parr and Russell(1997)}]{Parr1997ReinforcementLW}
Parr, R.~E.; and Russell, S.~J. 1997.
\newblock Reinforcement Learning with Hierarchies of Machines.
\newblock In \emph{NIPS}.

\bibitem[{Peng et~al.(2017)Peng, Wen, Yang, Yuan, Tang, Long, and
  Wang}]{Peng2017MultiagentBN}
Peng, P.; Wen, Y.; Yang, Y.; Yuan, Q.; Tang, Z.; Long, H.; and Wang, J. 2017.
\newblock Multiagent Bidirectionally-Coordinated Nets: Emergence of Human-level
  Coordination in Learning to Play StarCraft Combat Games.
\newblock \emph{arXiv: Artificial Intelligence}.

\bibitem[{Precup and Sutton(2000)}]{Precup2000TemporalAI}
Precup, D.; and Sutton, R. 2000.
\newblock Temporal abstraction in reinforcement learning.
\newblock In \emph{ICML 2000}.

\bibitem[{Rafati and Noelle(2019{\natexlab{a}})}]{Rafati2019LearningRI}
Rafati, J.; and Noelle, D.~C. 2019{\natexlab{a}}.
\newblock Learning Representations in Model-Free Hierarchical Reinforcement
  Learning.
\newblock In \emph{The Thirty-Third {AAAI} Conference on Artificial
  Intelligence, {AAAI} 2019, The Thirty-First Innovative Applications of
  Artificial Intelligence Conference, {IAAI} 2019, The Ninth {AAAI} Symposium
  on Educational Advances in Artificial Intelligence, {EAAI} 2019, Honolulu,
  Hawaii, USA, January 27 - February 1, 2019}, 10009--10010. {AAAI} Press.

\bibitem[{Rafati and Noelle(2019{\natexlab{b}})}]{Rafati2019UnsupervisedMF}
Rafati, J.; and Noelle, D.~C. 2019{\natexlab{b}}.
\newblock Unsupervised Methods For Subgoal Discovery During Intrinsic
  Motivation in Model-Free Hierarchical Reinforcement Learning.
\newblock In \emph{KEG@AAAI}.

\bibitem[{Rashid et~al.(2020)Rashid, Farquhar, Peng, and
  Whiteson}]{Rashid2020WeightedQE}
Rashid, T.; Farquhar, G.; Peng, B.; and Whiteson, S. 2020.
\newblock Weighted QMIX: Expanding Monotonic Value Function Factorisation for
  Deep Multi-Agent Reinforcement Learning.
\newblock \emph{arXiv: Learning}.

\bibitem[{Rashid et~al.(2018)Rashid, Samvelyan, de~Witt, Farquhar, Foerster,
  and Whiteson}]{Rashid2018QMIXMV}
Rashid, T.; Samvelyan, M.; de~Witt, C.~S.; Farquhar, G.; Foerster, J.~N.; and
  Whiteson, S. 2018.
\newblock {QMIX:} Monotonic Value Function Factorisation for Deep Multi-Agent
  Reinforcement Learning.
\newblock In Dy, J.~G.; and Krause, A., eds., \emph{Proceedings of the 35th
  International Conference on Machine Learning, {ICML} 2018,
  Stockholmsm{\"{a}}ssan, Stockholm, Sweden, July 10-15, 2018}, volume~80 of
  \emph{Proceedings of Machine Learning Research}, 4292--4301. {PMLR}.

\bibitem[{Samvelyan et~al.(2019)Samvelyan, Rashid, Witt, Farquhar, Nardelli,
  Rudner, Hung, Torr, Foerster, and Whiteson}]{Samvelyan2019TheSM}
Samvelyan, M.; Rashid, T.; Witt, C. S.~D.; Farquhar, G.; Nardelli, N.; Rudner,
  T. G.~J.; Hung, C.-M.; Torr, P. H.~S.; Foerster, J.~N.; and Whiteson, S.
  2019.
\newblock The StarCraft Multi-Agent Challenge.
\newblock In \emph{AAMAS}.

\bibitem[{Son et~al.(2019)Son, Kim, Kang, Hostallero, and Yi}]{Son2019QTRANLT}
Son, K.; Kim, D.; Kang, W.~J.; Hostallero, D.; and Yi, Y. 2019.
\newblock {QTRAN:} Learning to Factorize with Transformation for Cooperative
  Multi-Agent Reinforcement Learning.
\newblock In Chaudhuri, K.; and Salakhutdinov, R., eds., \emph{Proceedings of
  the 36th International Conference on Machine Learning, {ICML} 2019, 9-15 June
  2019, Long Beach, California, {USA}}, volume~97 of \emph{Proceedings of
  Machine Learning Research}, 5887--5896. {PMLR}.

\bibitem[{Song et~al.(2019)Song, Weng, Su, Yan, Zou, and
  Zhu}]{Song2019PlayingFG}
Song, S.; Weng, J.; Su, H.; Yan, D.; Zou, H.; and Zhu, J. 2019.
\newblock Playing {FPS} Games With Environment-Aware Hierarchical Reinforcement
  Learning.
\newblock In Kraus, S., ed., \emph{Proceedings of the Twenty-Eighth
  International Joint Conference on Artificial Intelligence, {IJCAI} 2019,
  Macao, China, August 10-16, 2019}, 3475--3482. ijcai.org.

\bibitem[{Su, Adams, and Beling(2021)}]{Su2021ValueDecompositionMA}
Su, J.; Adams, S.~C.; and Beling, P. 2021.
\newblock Value-Decomposition Multi-Agent Actor-Critics.
\newblock In \emph{AAAI}.

\bibitem[{Sukhbaatar, Szlam, and Fergus(2016)}]{Sukhbaatar2016LearningMC}
Sukhbaatar, S.; Szlam, A.; and Fergus, R. 2016.
\newblock Learning Multiagent Communication with Backpropagation.
\newblock In Lee, D.~D.; Sugiyama, M.; von Luxburg, U.; Guyon, I.; and Garnett,
  R., eds., \emph{Advances in Neural Information Processing Systems 29: Annual
  Conference on Neural Information Processing Systems 2016, December 5-10,
  2016, Barcelona, Spain}, 2244--2252.

\bibitem[{Sunehag et~al.(2018)Sunehag, Lever, Gruslys, Czarnecki, Zambaldi,
  Jaderberg, Lanctot, Sonnerat, Leibo, Tuyls, and
  Graepel}]{Sunehag2018ValueDecompositionNF}
Sunehag, P.; Lever, G.; Gruslys, A.; Czarnecki, W.; Zambaldi, V.; Jaderberg,
  M.; Lanctot, M.; Sonnerat, N.; Leibo, J.~Z.; Tuyls, K.; and Graepel, T. 2018.
\newblock Value-Decomposition Networks For Cooperative Multi-Agent Learning.
\newblock \emph{ArXiv}, abs/1706.05296.

\bibitem[{Sutton, Precup, and Singh(1999)}]{Sutton1999BetweenMA}
Sutton, R.; Precup, D.; and Singh, S. 1999.
\newblock Between MDPs and Semi-MDPs: A Framework for Temporal Abstraction in
  Reinforcement Learning.
\newblock \emph{Artif. Intell.}, 112: 181--211.

\bibitem[{Tang et~al.(2018)Tang, Hao, Lv, Chen, Zhang, Jia, Ren, Zheng, Meng,
  Fan, and Wang}]{Tang2018HierarchicalDM}
Tang, H.; Hao, J.; Lv, T.; Chen, Y.; Zhang, Z.; Jia, H.; Ren, C.; Zheng, Y.;
  Meng, Z.; Fan, C.; and Wang, L. 2018.
\newblock Hierarchical Deep Multiagent Reinforcement Learning with Temporal
  Abstraction.
\newblock \emph{arXiv: Learning}.

\bibitem[{van~der Maaten and Hinton(2008)}]{Maaten2008VisualizingDU}
van~der Maaten, L.; and Hinton, G.~E. 2008.
\newblock Visualizing Data using t-SNE.
\newblock \emph{Journal of Machine Learning Research}, 9: 2579--2605.

\bibitem[{Vezhnevets et~al.(2017)Vezhnevets, Osindero, Schaul, Heess,
  Jaderberg, Silver, and Kavukcuoglu}]{Vezhnevets2017FeUdalNF}
Vezhnevets, A.~S.; Osindero, S.; Schaul, T.; Heess, N.; Jaderberg, M.; Silver,
  D.; and Kavukcuoglu, K. 2017.
\newblock FeUdal Networks for Hierarchical Reinforcement Learning.
\newblock In Precup, D.; and Teh, Y.~W., eds., \emph{Proceedings of the 34th
  International Conference on Machine Learning, {ICML} 2017, Sydney, NSW,
  Australia, 6-11 August 2017}, volume~70 of \emph{Proceedings of Machine
  Learning Research}, 3540--3549. {PMLR}.

\bibitem[{Wang et~al.(2021)Wang, Gupta, Mahajan, Peng, Whiteson, and
  Zhang}]{Wang2021RODELR}
Wang, T.; Gupta, T.; Mahajan, A.; Peng, B.; Whiteson, S.; and Zhang, C. 2021.
\newblock RODE: Learning Roles to Decompose Multi-Agent Tasks.
\newblock \emph{ArXiv}, abs/2010.01523.

\bibitem[{Wiering and Hasselt(2009)}]{Wiering2009TheQF}
Wiering, M.~A.; and Hasselt, H.~V. 2009.
\newblock The QV family compared to other reinforcement learning algorithms.
\newblock \emph{2009 IEEE Symposium on Adaptive Dynamic Programming and
  Reinforcement Learning}, 101--108.

\bibitem[{Yang, Borovikov, and Zha(2020)}]{Yang2020HierarchicalCM}
Yang, J.; Borovikov, I.; and Zha, H. 2020.
\newblock Hierarchical Cooperative Multi-Agent Reinforcement Learning with
  Skill Discovery.
\newblock In \emph{AAMAS}.

\bibitem[{Zhang, Yu, and Xu(2021)}]{Zhang2021HierarchicalRL}
Zhang, J.; Yu, H.; and Xu, W. 2021.
\newblock Hierarchical Reinforcement Learning By Discovering Intrinsic Options.
\newblock \emph{ArXiv}, abs/2101.06521.

\end{thebibliography}





\onecolumn
\appendix
\numberwithin{equation}{section}
\numberwithin{figure}{section}
\numberwithin{table}{section}
\renewcommand{\thesection}{{\Alph{section}}}
\renewcommand{\thesubsection}{\Alph{section}.\arabic{subsection}}
\renewcommand{\thesubsubsection}{\Roman{section}.\arabic{subsection}.\arabic{subsubsection}}
\setcounter{secnumdepth}{-1}
\setcounter{secnumdepth}{3}

\section{Derivation of the Expected Start Value of the Low-Level Policies}
\label{sec:proof}

We give out the proof of the optimization target of low-level policies $\boldsymbol{\pi}^l$ as below:
\begin{align}
\eta\left(\tilde{\boldsymbol{\pi}}^{l}\right)
&=\mathbb{E}_{s_{0}^{l}}\left[V^{l}\left(s_{0}^{l}\right)\right]\nonumber\\
&=\mathbb{E}_{\tau \sim\left(\tilde{\boldsymbol{\pi}}^{l}, \boldsymbol{\pi}^{h}\right)}\left[\sum_{t=0,1,2, \ldots} \gamma_{l}^{t} r\left(s_{t}^{l}, \boldsymbol{u}_{t}^{l}\right)\right]\nonumber\\
&=\mathbb{E}_{\tau \sim\left(\tilde{\boldsymbol{\pi}}^{l}, \boldsymbol{\pi}^{h}\right)}\left[\sum_{t=0, k, 2 k, \ldots} \mathbb{E}_{\tau_{l}(t) \sim\left(\tilde{\boldsymbol{\pi}}^{l}, \boldsymbol{\pi}^{h}\right)}\left[\sum_{i=0}^{k-1} \gamma_{l}^{t+i} \left(\frac{1}{k}A_{h}\left(s_{t}^{h}, \boldsymbol{u}_{t}^{h}\right)+r^e\left(s_{t+i}^l,\boldsymbol{u}^l_{t+i}\right)\right)\right]\right]\label{eq:a.1}\\
&= \mathbb{E}_{\tau_{h} \sim\left(\tilde{\boldsymbol{\pi}}^{l}, \boldsymbol{\pi}^{h}\right)}\left[\sum_{T=0, 1, 2, \ldots} \sum_{i=0}^{k-1} \gamma_{l}^{T\cdot k+i} \left(\frac{1}{k}A_{h}\left(s_{T}^{h}, \boldsymbol{u}_{T}^{h}\right)\right)\right] \nonumber\\
& \qquad+ \mathbb{E}_{\tau_{h} \sim\left(\tilde{\boldsymbol{\pi}}^{l}, \boldsymbol{\pi}^{h}\right)}\left[\sum_{t=0, k, 2 k, \ldots} \mathbb{E}_{\tau_{l}(t) \sim\left(\tilde{\boldsymbol{\pi}}^{l}, \boldsymbol{\pi}^{h}\right)}\left[\sum_{i=0}^{k-1} \gamma_{l}^{t+i} r^e\left(s^l_{t+i},\boldsymbol{u}^l_{t+i}\right)\right]\right]\label{eq:a.2}\\
&\approx\frac{1}{k}\mathbb{E}_{\tau_{h} \sim\left(\tilde{\boldsymbol{\pi}}^{l}, \boldsymbol{\pi}^{h}\right)}\left[\sum_{T=0, 1, 2, \ldots} \gamma_{l}^{T\cdot k} \frac{1-\gamma_{l}^{k}}{1-\gamma_{l}} A_{h}\left(s_{T}^{h}, \boldsymbol{u}_{T}^{h}\right)\right]+\mathbb{E}_{\tau_{h} \sim\left(\tilde{\boldsymbol{\pi}}^{l}, \boldsymbol{\pi}^{h}\right)}\left[\sum_{T=0, 1, 2, \ldots} \gamma^T_h R\left(s^h_T,\boldsymbol{u}^h_T\right)\right]\label{eq:a.3}\\
&\approx \frac{1-\gamma_{l}^{k}}{k(1-\gamma_{l})} \mathbb{E}_{\tau_{h} \sim\left(\tilde{\boldsymbol{\pi}}^{l}, \boldsymbol{\pi}^{h}\right)}\left[\sum_{T=0, 1, 2, \ldots} \gamma_{l}^{T\cdot k} A_{h}\left(s_{T}^{h}, \boldsymbol{u}_{T}^{h}\right)\right]+\eta(\tilde{\boldsymbol{\pi}}^{joint})\label{eq:a.4}\\
&\approx \frac{1-\gamma_{l}^{k}}{k(1-\gamma_{l})} \mathbb{E}_{\tau_{h} \sim\left(\tilde{\boldsymbol{\pi}}^{l}, \boldsymbol{\pi}^{h}\right)}\left[\sum_{T=0, 1, 2, \ldots} \gamma_{h}^{T} A_{h}\left(s_{T}^{h}, \boldsymbol{u}_{T}^{h}\right)\right]+\eta(\boldsymbol{\pi}^{joint})\nonumber\\
&\qquad+ \mathbb{E}_{\tau_{h} \sim\left(\tilde{\boldsymbol{\pi}}^{l}, \boldsymbol{\pi}^{h}\right)}\left[\sum_{T=0, 1, 2 , \ldots} \gamma_{h}^{T} A_{h}\left(s_{T}^{h}, \boldsymbol{u}_{T}^{h}\right)\right]\\
&=\left[1+\frac{1-\gamma_{l}^{k}}{k(1-\gamma_{l})}\right]\mathbb{E}_{\tau_{h} \sim\left(\tilde{\boldsymbol{\pi}}^{l}, \boldsymbol{\pi}^{h}\right)}\left[\sum_{T=0, 1, 2 , \ldots} \gamma_{h}^{T} A_{h}\left(s_{T}^{h}, \boldsymbol{u}_{T}^{h}\right)\right] + \eta(\boldsymbol{\pi}^{joint})\nonumber
\end{align}
The last term in Eq.~\eqref{eq:a.1} is derived from the definition of the low-level reward function $r=r^i+r^e$. We regard $\sum_{t=0, k, 2 k, \ldots} \mathbb{E}_{\tau_{l}(t) \sim\left(\tilde{\boldsymbol{\pi}}^{l}, \boldsymbol{\pi}^{h}\right)}\left[\sum_{i=0}^{k-1} \gamma_{l}^{t+i} r^e\left(s^l_{t+i},\boldsymbol{u}^l_{t+i}\right)\right]$ and $\sum_{T=0, 1, 2, \ldots} \gamma^T_h R\left(s^h_T,\boldsymbol{u}^h_T\right)$ as equal under the condition of that $\gamma^l$ as well as $\gamma^h$ are both close to 1 and the $k$ is not extremely large. Then Eq.~\eqref{eq:a.2} can be written as Eq.~\eqref{eq:a.3} by following the above assumption. Substituting Eq.~\eqref{eq:pi_target} into Eq.~\eqref{eq:a.3} yields Eq.~\eqref{eq:a.4}, which is only with reference to $\boldsymbol{\pi}^h$. Finally, we replace the optimization target of updated joint policies $\tilde{\boldsymbol{\pi}}^{joint}$ with Eq.~\eqref{eq:tilde_pi_target} and get the last form of the optimization target of $\boldsymbol{\pi}^l$.

\section{Implementation Details}

\subsection{Hyperparameters}
\label{sec:app_hyper}
Hyperparameters were based on the PyMARL implementation and are listed in Table~\ref{table:hyperparameters}. All experiments in this paper are run on Nvidia GeForce RTX 3090 graphics cards and Intel(R) Xeon(R) Platinum 8280 CPU. The epsilon annealing period (for epsilon-greedy exploration) is 50000 steps and in order to be fair, we set the $\epsilon$ of the high-level policy to be consistent with that of the low-level policy. 

\begin{table}[htbp]
\centering
\begin{tabular}{lll}
\hline
\textbf{Name}                                                                      & \textbf{Description}                                & \textbf{Value} \\ \hline
                                                                                   & Learning rate                                       & 0.0005         \\
                                                                                   & Type of optimizer                                   & RMSProp        \\
optim $\alpha$                                                                     & RMSProp param                                       & 0.99           \\
optim $\epsilon$                                                                   & RMSProp param                                       & 0.00001        \\
                                                                                   & How many episodes to update target networks         & 200            \\
                                                                                   & Reduce global norm of gradients                     & 10             \\
                                                                                   & Batch size                                          & 32             \\
                                                                                   & Capacity of replay buffer (in episodes)             & 5000           \\
$\gamma_h$, $\gamma_l$                                                             & Discount factor                                     & 0.99           \\
starting $\epsilon$                                                                & Starting value for exploraton rate annealing        & 1              \\
ending $\epsilon$                                                                  & Ending value for exploraton rate annealing          & 0.05           \\ \hline
$k$                                                                                & How many timesteps to execute the high-level policy & 3              \\
$N$                                                                                & Number of macro actions                             & 8              \\ \hline
\end{tabular}
\caption{Hyperparameter settings.}
\label{table:hyperparameters}
\end{table}

\subsection{Algorithmic Description}

The algorithm for HAVEN are summarized in Algorithm 1. The code for HAVEN can be found in the supplementary material.

\begin{algorithm}[htbp]
\caption{HAVEN}
\label{alg:COLA-QMIX}
\textbf{Hyperparameters}: the high-level time scale $k$, the number of macro actions $N$, the discount factor $\gamma$\\
Initialize the parameters of the agent network and the mixing network\\
Initialize the parameters of the macro agent network and the macro mixing network\\
Initialize the parameters of the value agent network and the value mixing network\\
Initialize the low-level replay buffer $\mathcal{D}^l$\\
Initialize the high-level replay buffer $\mathcal{D}^h$
\begin{algorithmic}[1] 
\FOR{each episode}
\STATE Obtain the global state $s_0$ and the local observations $\boldsymbol{z}_0=\{z^1_0, z^2_0,\dots, z^n_0\}$
\FOR{$t \leftarrow 0$ to $\infty$}
\FOR{$a \leftarrow 1$ to $n$}
\IF{$t \mod k == 0$}
\STATE Select macro action $u^{h,a}_{t/k}$ according to $\epsilon$-greedy high-level policy w.r.t $Q_a^h(\tau^{h,a}_{t/k}, \cdot)$
\ENDIF
\STATE Select primitive action $u^{l,a}_t$ according to $\epsilon$-greedy low-level policy w.r.t $Q_a^l(\tau^{l,a}_t, u^{h,a}_{\lfloor t/k\rfloor}, \cdot)$
\ENDFOR
\STATE Take the joint action $\boldsymbol{u}^l_t=\{u_t^{l, 1}, u_t^{l, 2},\dots,u_t^{l, n}\}$
\STATE Obtain the external reward $r_{t}^e$, the next local observations $\boldsymbol{z}_{t+1}$, and the next state $s_{t+1}$
\IF{$t \mod k == 0 \ \text{and}\ t > 0$}
\STATE Calculate the high-level reward $R_{t/k-1}$
\STATE Store $\langle s_{t/k-1},\boldsymbol{z}_{t/k-1},\boldsymbol{u}^h_{t/k-1}, R_{t/k-1} \rangle$ in $\mathcal{D}^h$
\ENDIF
\STATE Store $\langle s_t,\boldsymbol{z}_t, \boldsymbol{u}^h_{\lfloor t/k\rfloor},\boldsymbol{u}^l_t, r^e_t\rangle$ in $\mathcal{D}^l$
\ENDFOR
\STATE Sample a high-level batch of episodes $\mathcal{B}^h \sim$ Uniform($\mathcal{D}^h$)
\STATE Update the parameters of the value mixing network and the value agent network according Eq.~(8)
\STATE Update the parameters of the macro mixing network and the macro agent network according Eq.~(9)
\STATE Sample a low-level batch of episodes $\mathcal{B}^l \sim$ Uniform($\mathcal{D}^l$)
\STATE Calculate the intrinsic reward $r^i$ for $\mathcal{B}^l$ according Eq.~(1)
\STATE Update the parameters of the agent network and the mixing network according Eq.~(10)
\STATE Replace target parameters every $M$ episodes
\ENDFOR
\end{algorithmic}
\end{algorithm}

\newpage


\section{Experiment Details}
\label{sec:app_experiment}

\begin{figure*}[ht]
    \centering
    \subfigure[SMAC]{
        \includegraphics[width=1.8 in]{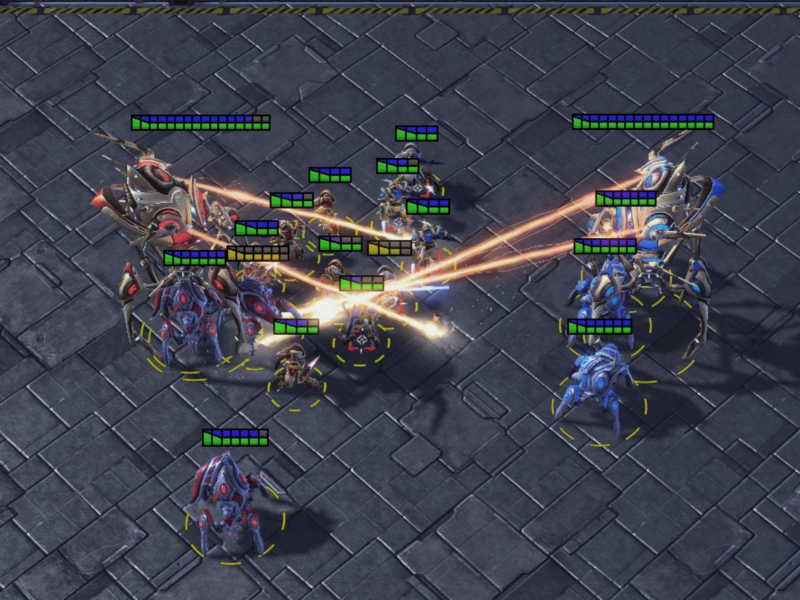}
        \label{fig:sc2}
    }
    \hspace{0.7 in}
    \subfigure[Google Research Football]{
        \includegraphics[width=1.8 in]{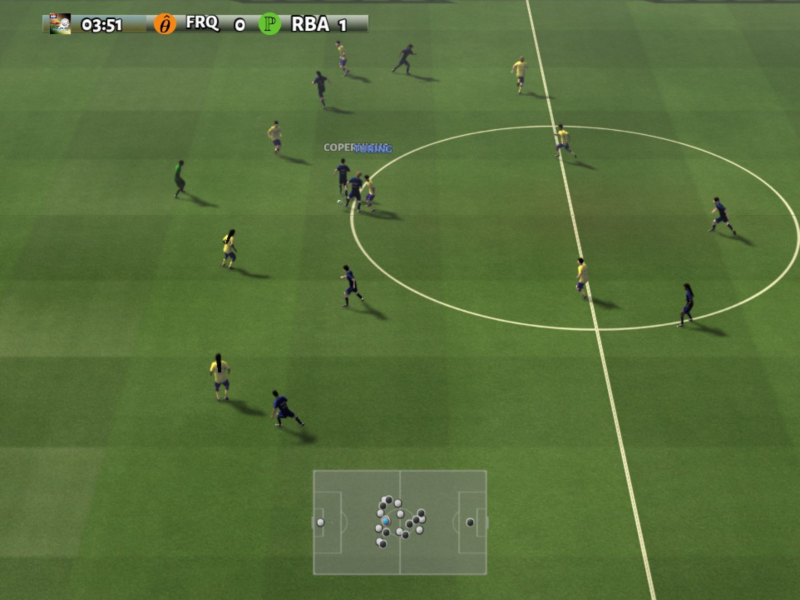}
        \label{fig:grf}
    }
    \vskip -0.15 in
    \caption{Screenshots of the two experimental platforms used in this paper.}
\end{figure*}

\subsection{StarCraft II Micromanagement Tasks}
\label{sec:app_sc2}

Depending on the complexity of the scenarios, the duration of each experiment ranges from 6 to 14 hours. The detailed information of all scenarios is summarized in Table~\ref{table:scenario}. We run all experiments independently for evaluation with five different random seeds. Since the difficulty of the map is determined based on version 4.6.2.69232 of StarCraft II, we carried out all experiments on StarCraft II of this relatively more difficult version instead of version 4.10. The results of some benchmarks in this paper are different from those in the literature, which may be caused by inconsistent versions of StarCraft II.
\begin{table}[h]
\centering
\begin{tabular}{lcccc}
\hline
\textbf{Name}  & \textbf{Ally Units}                                                         & \textbf{Enemy Units}                                                        & \textbf{Type}                                                                         & \textbf{Difficulty} \\ \hline
2s3z           & \begin{tabular}[c]{@{}c@{}}2 Stalkers\\ 3 Zealots\end{tabular}              & \begin{tabular}[c]{@{}c@{}}2 Stalkers\\ 3 Zealots\end{tabular}              & \begin{tabular}[c]{@{}c@{}}Heterogeneous\\ Symmetric\end{tabular}                     & Easy       \\ \hline
3s5z           & \begin{tabular}[c]{@{}c@{}}3 Stalkers\\ 5 Zealots\end{tabular}              & \begin{tabular}[c]{@{}c@{}}3 Stalkers\\ 5 Zealots\end{tabular}              & \begin{tabular}[c]{@{}c@{}}Heterogeneous\\ Symmetric\end{tabular}                     & Easy       \\ \hline
1c3s5z         & \begin{tabular}[c]{@{}c@{}}1 Colossus\\ 3 Stalkers\\ 5 Zealots\end{tabular} & \begin{tabular}[c]{@{}c@{}}1 Colossus\\ 3 Stalkers\\ 5 Zealots\end{tabular} & \begin{tabular}[c]{@{}c@{}}Heterogeneous\\ Symmetric\end{tabular}                     & Easy       \\ \hline
5m\_vs\_6m     & 5 Marines                                                                   & 6 Marines                                                                   & \begin{tabular}[c]{@{}c@{}}Homogeneous\\ Asymmetric\end{tabular}                      & hard       \\ \hline
3s\_vs\_5z     & 3 Stalkers                                                                  & 5 Zealots                                                                   & \begin{tabular}[c]{@{}c@{}}Homogeneous\\ Asymmetric\end{tabular}                      & hard       \\ \hline
bane\_vs\_bane & \begin{tabular}[c]{@{}c@{}}4 Banelings\\ 20 Zerglings\end{tabular}          & \begin{tabular}[c]{@{}c@{}}4 Banelings\\ 20 Zerglings\end{tabular}          & \begin{tabular}[c]{@{}c@{}}Heterogeneous\\ Symmetric\end{tabular}                     & hard       \\ \hline
2c\_vs\_64zg   & 2 Colossi                                                                   & 64 Zerglings                                                                & \begin{tabular}[c]{@{}c@{}}Homogeneous\\ Asymmetric\\ Large Action Space\end{tabular} & hard       \\ \hline
MMM2           & \begin{tabular}[c]{@{}c@{}}1 Medivac\\ 2 Marauders\\ 7 Marines\end{tabular} & \begin{tabular}[c]{@{}c@{}}1 Medivac\\ 3 Marauder\\ 8 Marines\end{tabular}  & \begin{tabular}[c]{@{}c@{}}Heterogeneous\\ Asymmetric\\ Macro tactics\end{tabular}    & Super Hard \\ \hline
27m\_vs\_30m   & 27 Marines                                                                  & 30 Marines                                                                  & \begin{tabular}[c]{@{}c@{}}Homogeneous\\ Asymmetric\\ Massive Agents\end{tabular}     & Super Hard \\ \hline
\end{tabular}
\caption{Maps in different scenarios.}
\label{table:scenario}
\end{table}

\subsection{Google Research Football Tasks}

\begin{figure*}[ht]
    \centering
    \subfigure[]{
        \includegraphics[width=1.3 in]{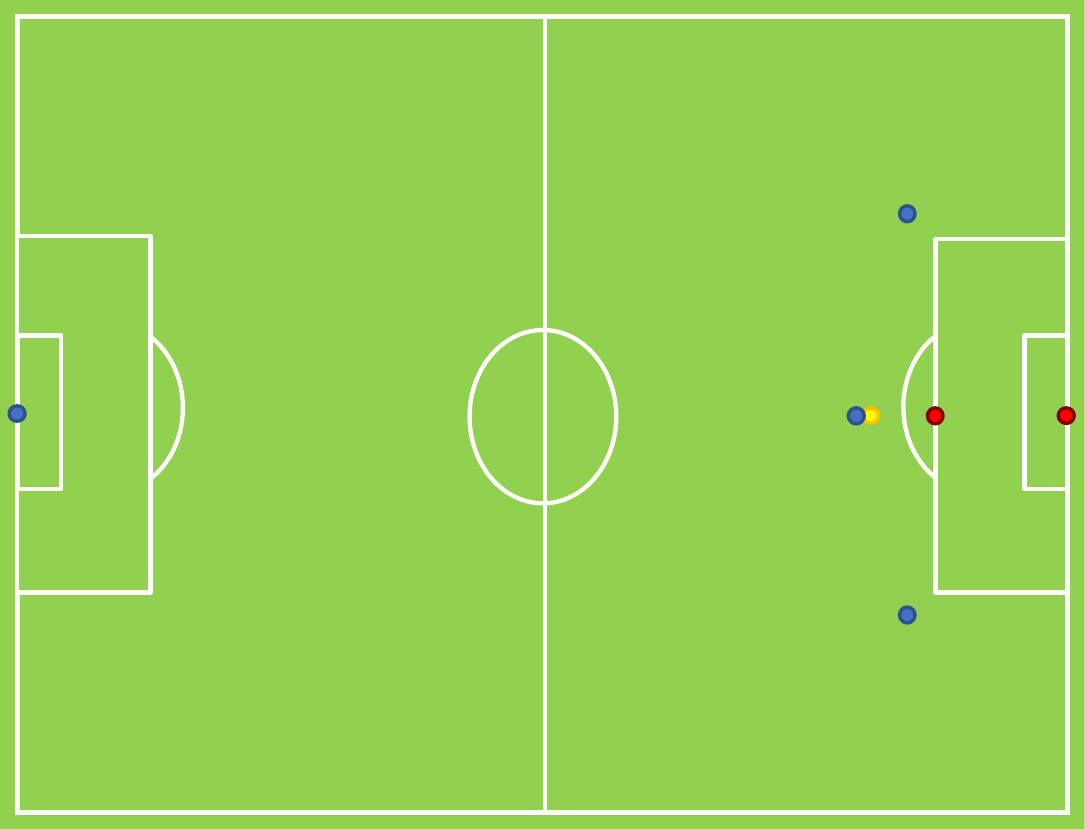}
        \label{fig:academy_3_vs_1_with_keeper}
    }
    \subfigure[]{
        \includegraphics[width=1.3 in]{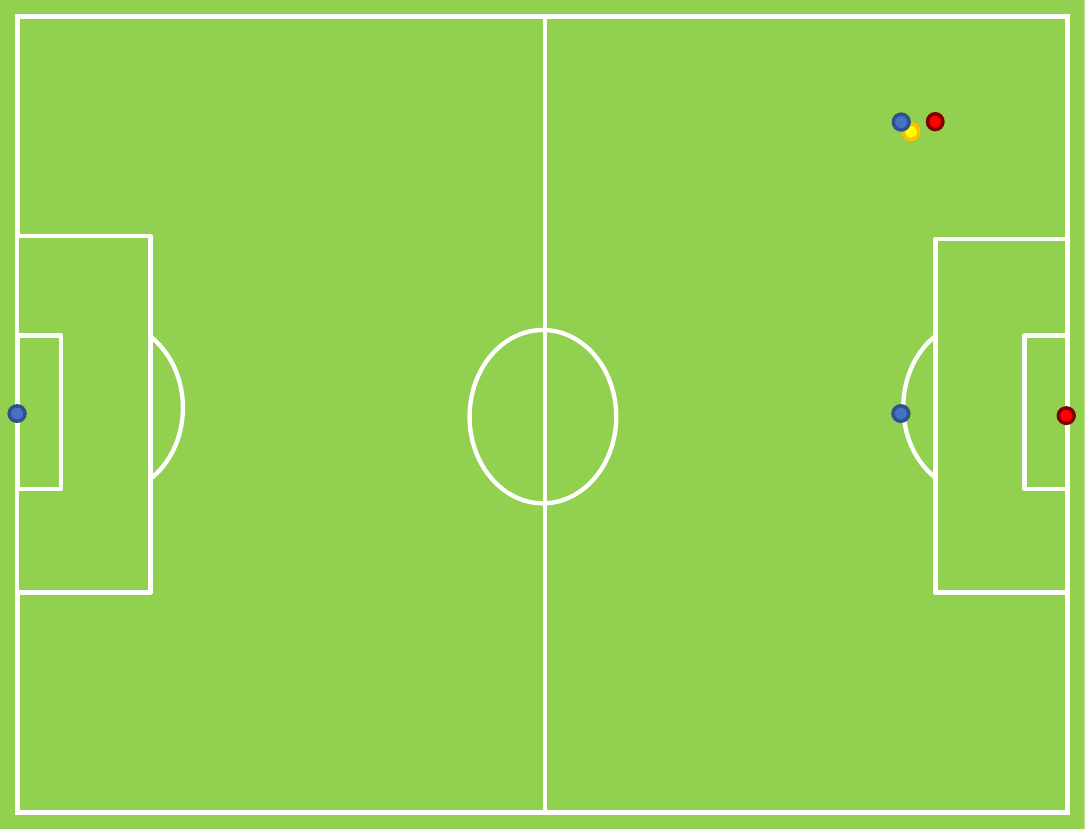}
        \label{fig:academy_pass_and_shoot_with_keeper}
    }
    \subfigure[]{
        \includegraphics[width=1.3 in]{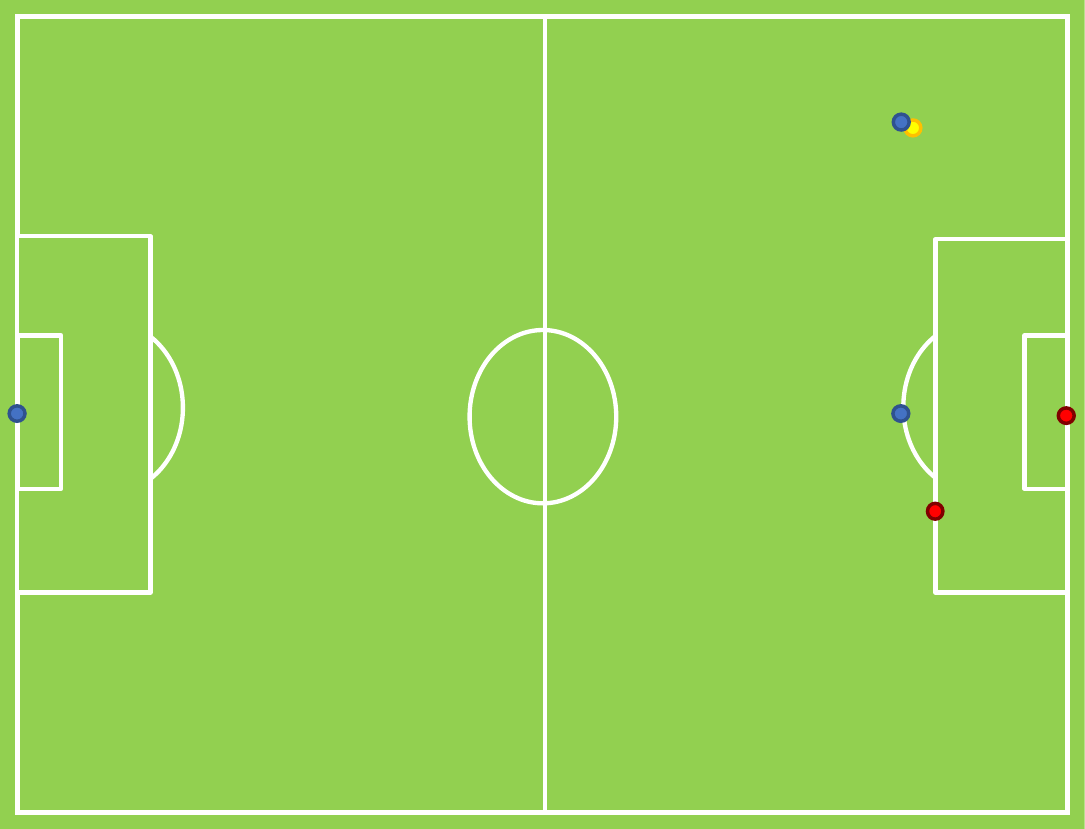}
        \label{fig:academy_run_pass_and_shoot_with_keeper}
    }
    \subfigure[]{
        \includegraphics[width=1.3 in]{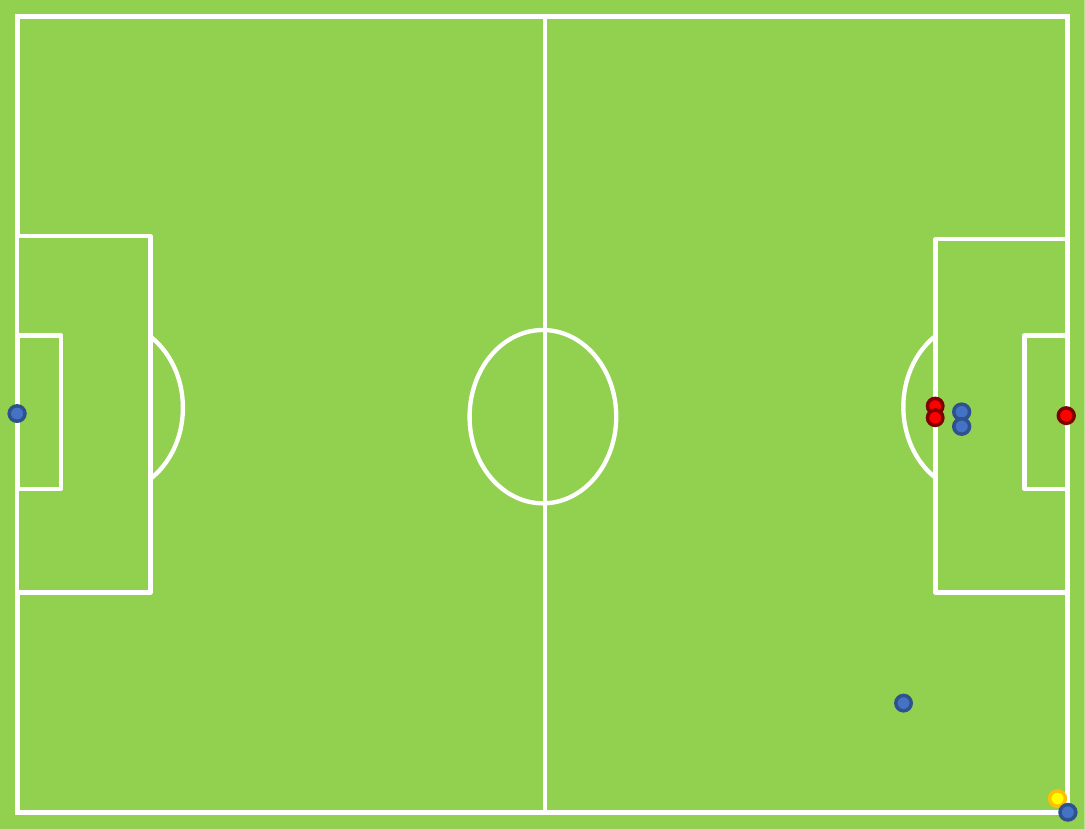}
        \label{fig:academy_corner}
    }
    \vskip -0.15 in
    \caption{The initial position of each agent in the Google Research Football environments considered in our paper. (a) \emph{academy\_3\_vs\_1\_with\_keeper}. (b) \emph{academy\_pass\_and\_shoot\_with\_keeper}. (c) \emph{academy\_run\_pass\_and\_shoot\_with\_keeper}. (d) \emph{academy\_corner}. The blue and red points denote our players and the opposing players, respectively. The ball is represented by the yellow point.}
    \label{fig:grf_maps}
\end{figure*}

The agents in the Google Research Football environment have 19 actions, including standard movement actions as well as different ways to kick the ball, such as passes and shooting. We employ a shaped reward, which augments the scoring reward with an additional auxiliary reward contribution for moving the ball close to the opponent's goal. To investigate the robustness of HAVEN, we run the environment in stochastic mode, which means that several types of randomness are introduced into the transition function. To speed up training, we set that the episode is judged to be terminated when the ball returns to our half or enters the opponent's goal. The settings of the four official maps we used in the paper are shown in Figure~\ref{fig:grf_maps}. All experiments on the Google Research Football environments were done in two days.

\section{Additional Visualization}

\begin{figure*}[h]
    \centering
    \subfigure[High-level policies.]{
        \includegraphics[width=2.3 in]{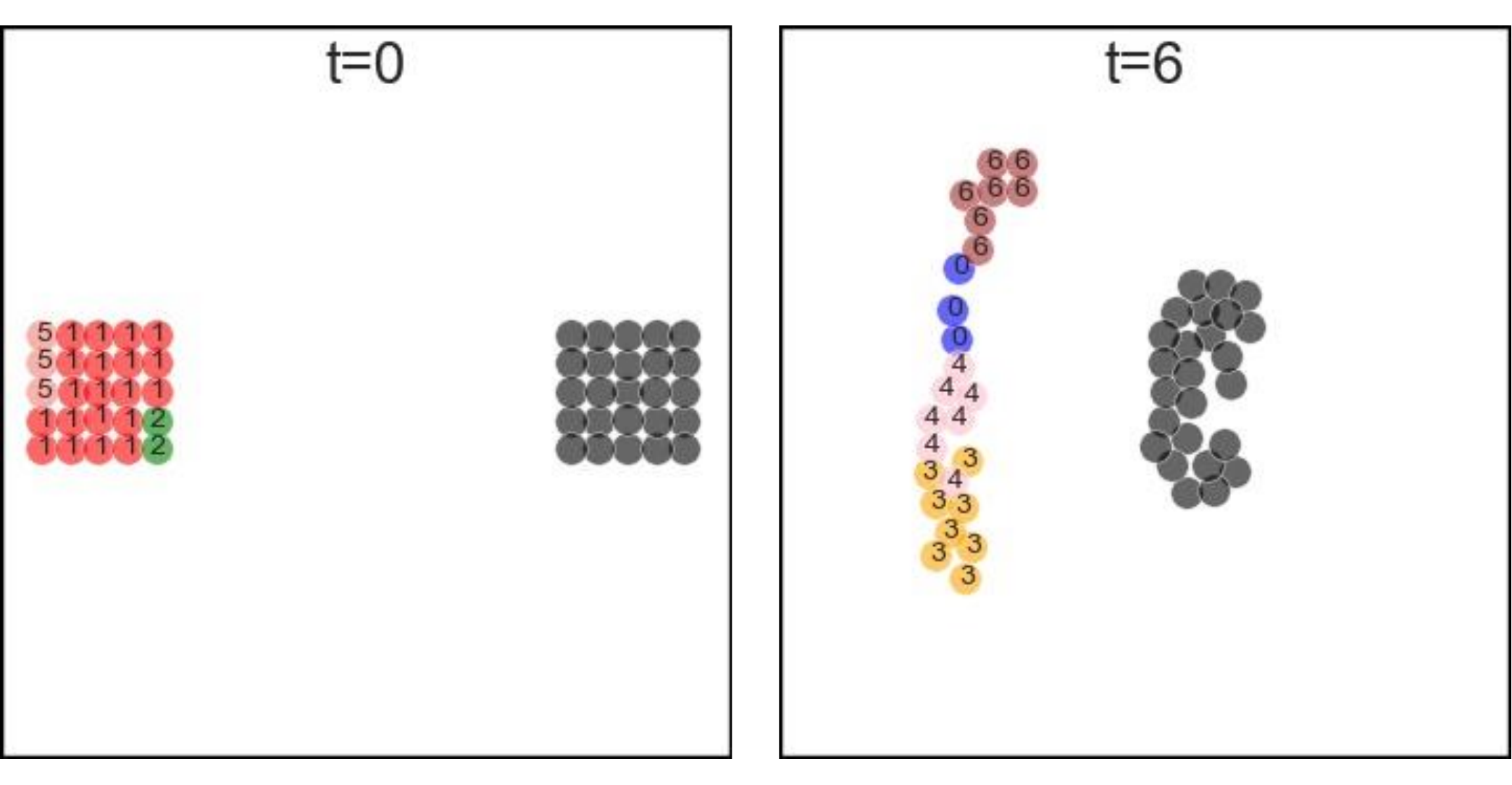}
        \label{fig:high_level_pos_action}
    }
    \hspace{0.5 in}
    \subfigure[Low-level policies.]{
        \includegraphics[width=2.3 in]{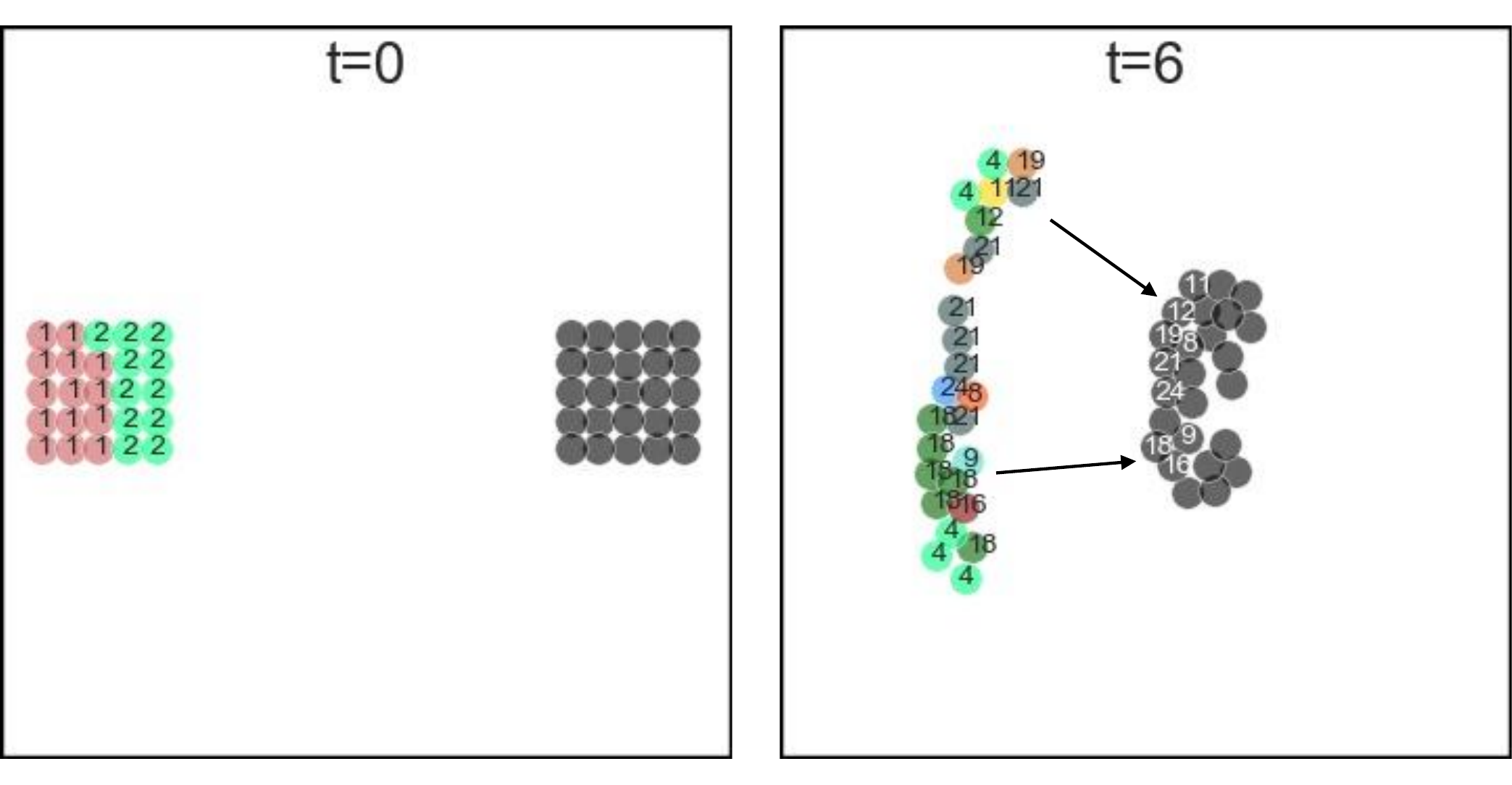}
        \label{fig:low_level_pos_action}
    }
    \vskip -0.15 in
    \caption{Two game snapshots in one episode on $25m$ scenario. Gray points are the enemies and color points are allies. The different colors correspond to different actions. Note that the action spaces at different levels are different. In the low-level action space, actions 1-5 are movement-related actions, and actions 6-30 are attack-related actions. We labeled some enemy units with serial numbers corresponding to low-level attack-related actions.}
    \label{fig:snapshot}
\end{figure*}

The coordination of intra-level policies is depicted in Figure~\ref{fig:snapshot}. In Figure~D.\ref{fig:high_level_pos_action}, we depict a one-to-one scatter diagram of the positions and macro actions at two different timesteps in one episode on scenario \emph{25m}, presenting a massive multi-agent task. Neighbour agents often choose the same macro actions, consistent with the practical multi-agent system. Similarly, from the perspective of low-level policies, the adjacent agents in Figure~D.\ref{fig:low_level_pos_action} tend to move in one direction or attack enemies in the same area.

\begin{figure*}[ht]
    \centering
    \includegraphics[width=4.0 in]{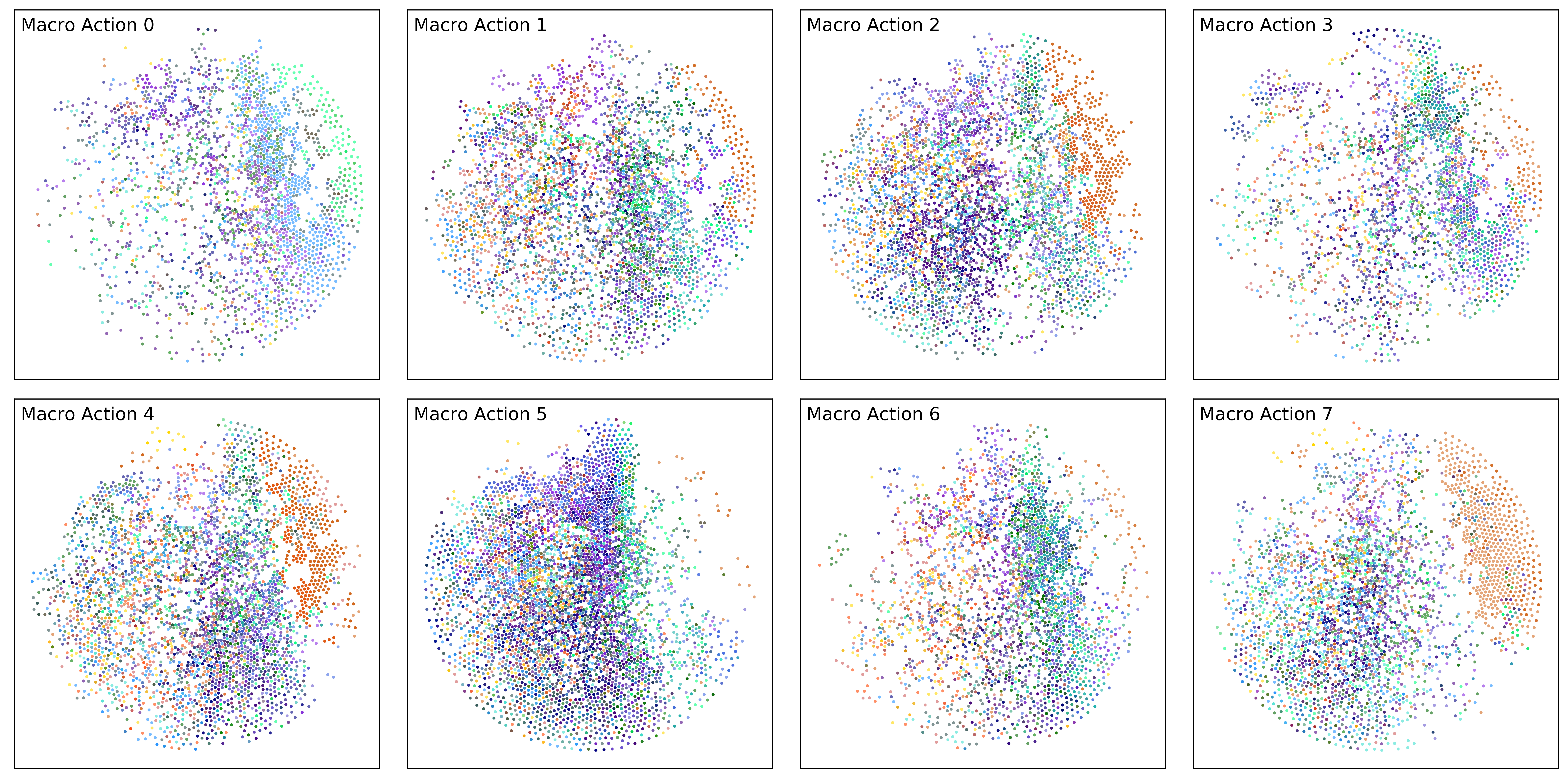}
    \vskip -0.15 in
    \caption{The 2D embeddings of the decision spaces corresponding to all macro actions.}
    \label{fig:add_visual}
\end{figure*}

As shown in Figure~\ref{fig:add_visual}, we show the 2D embedding of the decision space corresponding to all macro actions. Blank areas indicate that agents will not select the corresponding macro action in these states, meaning high-level policies can automatically partition the decision space without domain knowledge. Each macro action is only responsible for decisions in a certain state subspace. Through the hierarchical structure, HAVEN significantly reduces the difficulty of the exploration and selection in the large action space.

\end{document}